%% file: arXiv_mpls_sdn.tex
\documentclass[journal,12pt,onecolumn,draftclsnofoot]{IEEEtran}

\usepackage[T1]{fontenc}
\usepackage{etex}
\usepackage{booktabs}
\usepackage[caption=false,font=footnotesize]{subfig}
\usepackage{graphicx}
\usepackage{import, color}
\usepackage{verbatim}
\usepackage{soul}
\usepackage{epstopdf}
\usepackage{amssymb}
\usepackage{listings}
\usepackage{color}
\usepackage{tabularx}
\usepackage{mathtools}
\usepackage{multirow}
\usepackage{booktabs}
\usepackage{wrapfig}
\usepackage{amsmath}
\usepackage{listings}
\usepackage{framed}
\usepackage{paralist}
\usepackage{todonotes}
\usepackage{bigstrut}
\usepackage{pdflscape}
\usepackage{rotating}

\definecolor{dkgreen}{rgb}{0,0.6,0}
\definecolor{gray}{rgb}{0.5,0.5,0.5}
\definecolor{mauve}{rgb}{0.58,0,0.82}

\DeclareMathOperator*{\avg}{avg}

\makeatletter
\def\ps@pprintTitle{%
   \let\@oddhead\@empty
   \let\@evenhead\@empty
   \let\@oddfoot\@empty
   \let\@evenfoot\@oddfoot
}
\makeatother

\begin{document}

\title{MPLS-based Reduction of Flow Table Entries in SDN Switches Supporting Multipath Transmission}

\author{Zbigniew~Duli\'nski,~
        Grzegorz~Rzym,~
        and~Piotr~Cho{\l{}}da 
\thanks{Z.~Duli\'nski is with Jagiellonian University, Faculty of Physics, Astronomy and Applied Computer Science;
Ul.~\L{}ojasiewicza 11, 30-348 Krak\'ow, Poland, e-mail: dulinski@th.if.uj.edu.pl}
\thanks{G.~Rzym and P.~Cho{\l{}}da are with AGH University of Science and Technology, Department of Telecommunications; Al.~Mickiewicza 30, 30-059 Krak\'ow, Poland, e-mail: \{rzym,cholda\}@agh.edu.pl.}
}

\maketitle

\begin{abstract}        
In the paper, a new mechanism for Software-Defined Networking (SDN) flow aggregation accompanied with multipath transmission is proposed. The aggregation results in a low number of flow entries in core switches. This type of scalability improvement of flow processing is obtained thanks to application of the procedure based on introduction of centrally managed MPLS label distribution performed by an SDN controller. Moreover, multipath transmission improves network resource utilization.

The proposed mechanism does not involve definition of new protocols. It neither needs application of legacy signalling protocols. Only simple modifications of the existing solutions, assured by flexibility of OpenFlow, are necessary. Additionally, the mechanism significantly reduces communication between the controller and switches, thus enabling a fine-grained flow-level control. Furthermore, the proposed solution can be incrementally deployed in legacy networks. The simulations show that a number of flow entries in core switches can be reduced as much as of 96\%, while the overall network traffic is increased by about 32\%. Moreover, the number of OpenFlow messages is reduced by about 98\%.
\end{abstract}


\newpage
\input{sections/intro}

\input{sections/motivation}
\input{sections/mechanism}
\input{sections/evaluation}
\input{sections/related}
\input{sections/conclusion}
\bibliographystyle{IEEEtran}
\bibliography{./bibliography}

\end{document}

%% file: sections/intro.tex
\section{Introduction} \label{sec:intro}

In legacy networks, packets traverse the metric-related best single path between a pair of source and destination nodes. 
When a congestion appears on this path, its route should be changed.
However, a modification of even a single metric (weight) can be disruptive to a whole network due to scalability issues: %
\begin{inparaenum}[(a)]
	\item update of routing tables takes a considerable amount of time, and %
	\item it is likely to cause reordering or packet dropping, thus decreasing the performance of TCP~\cite{fortz2002}.
\end{inparaenum}
Obviously, the more changes are introduced, the larger chaos is observed.
On the other hand, the well known fact is that
in almost any network there is at least one concurrent path alternative to the one used. This fact enables to counteract the mentioned congestion problem with the so-called multipath transmission. Various multipath transmission methods can be performed in different network layers~\cite{multipath-survey1, multipath-survey2}.
Apart from using additional paths, this type of transmission assumes that the routing is semi-independent of current link weights. Nowadays, the most popular solution for establishing such paths is the Multiprotocol Label Switching (MPLS) enabling flexible traffic engineering~\cite{rfc3031}. However, MPLS paths are established for a long-time scale and large amounts of data. Therefore, although these paths can be periodically reoptimized, such a process again results in disruption of existing traffic and typically does not take into account present utilization of links.

Disruption-free transmission of packets using a path 
that can be changed with the fine granularity of time or data volumes is solved by introduction of the flow-based forwarding in Software-Defined Networking (SDN)~\cite{sdnsurvey, sdnwhitepaper}. 
Unfortunately, application of flow-based switches supporting fine-grained flow-level control
results in increase of unmanageable flow table sizes, providing scalability problems.
This fact hinders flow-based forwarding due to storage limitations and 
lookup delays. Such a problem has already been noticed with introduction of the OpenFlow protocol~\cite{openflow, openflowspec}. 
The issue has been addressed and, notably, Ternary Content Addressable Memory (TCAM) is used for storing flow entries~\cite{curtis2011,khalili2016,chuang2016}.
Moreover, a centralized management approach can create significant signaling overhead, especially when the \textit{reactive} mode for flow installation is used. Then, extensive communication between an SDN controller and switches is required~\cite{commag-scalability}. Early benchmarks show that controllers are unable to handle a huge number of requests~\cite{nox, controller-performance}. This problem is mostly burdening in Data Center (DC) environments, where enormous numbers of flows are present~\cite{benson, Kandula}. On the other hand, the \textit{proactive} mode can be advantageous, but such a solution is traded off with precision of traffic management. Another scalability problem is related to flow installation time in hardware switches. This time can reach sub-10~ms latency levels at best~\cite{commag-scalability}.

We propose to further reduce the size of flow tables in the core of a network, which supports multipath transmission. Simultaneously, we minimize signalling requirements. The mechanism is based on tagging flows with the MPLS approach, where forwarding of packets is performed on the basis of labels. However, any tagging mechanism (e.g., VLANs) could be used. In the proposed mechanism the distribution of labels is not supported by Label Distribution Protocol (LDP) or Resource Reservation Protocol (RSVP). Instead, we use OpenFlow. Also, Segment Routing \cite{sr-draft} can provide such a functionality. The proposed mechanism explores only the existing off-the-shelf solutions (MPLS, SDN, OSPF/IS-IS, LLDP) by originally combining the following two elements.
\begin{inparaenum}[(1)]
    \item \textit{Multipath transmission} enabling a better network resource utilization based on fast reaction to network condition changes. Thus, we ensure that a congestion appearing on a link when other links are underutilized is solved by on-demand and automated path recalculation.
    \item \textit{MPLS-based flow aggregation} enabling decrease of flow tables in switches centrally managed by an SDN controller. The forwarding decision on traffic flows destined to a selected node is based on a single label in a whole network.
\end{inparaenum}

The paper is organized as follows.
Section~\ref{sec:motivation} presents the justification behind introduction of the proposed mechanism.
Section~\ref{sec:mechanism} introduces the mechanism and its architecture, that is the solution for a centralized path set-up optimization supporting IP flows.
Section~\ref{sec:evaluation} describes the evaluation details, including the tools used and performance results. Also mechanism scalability is discussed.
Section~\ref{sec:related} contains a review of the related work together with a comparison of our approach and the ones presented before.
Section~\ref{sec:conclusion} summarizes the paper with concise conclusions. 

%% file: sections/motivation.tex
\section{Problem Statement and Motivation behind the Mechanism}\label{sec:motivation}

In this section, we briefly describe drivers for the proposed mechanism. This relates to three important problems appearing in networks operating with the flow-forwarding scheme, namely:
\begin{inparaenum}[(a)]
    \item scalability of flow tables at switches in the core of the network,
    \item link congestions, and
    \item flow installation overhead.
\end{inparaenum}

\subsection{Flow Aggregation}\label{subsec:aggregation}

Flow-based forwarding supports effective traffic distinction and management. However, this approach suffers from a huge number of flow entries that need to be maintained by each flow-forwarding node. It is a well-known fact that TCAM is the most suitable memory technology for flow storing and forwarding~\cite{sdnsurvey}. However, it is very expensive, consumes a lot of energy, and can store a limited number of entries only~\cite{tcam-razor,curtis2011,khalili2016,chuang2016,Katta2014}. The last drawback is most important from the viewpoint of TCAM applications. The number of entries which has to be served by a switch strongly depends on a level of network aggregation hierarchy.

\begin{figure}[ht]
    \centering
    \includegraphics[scale=0.57]{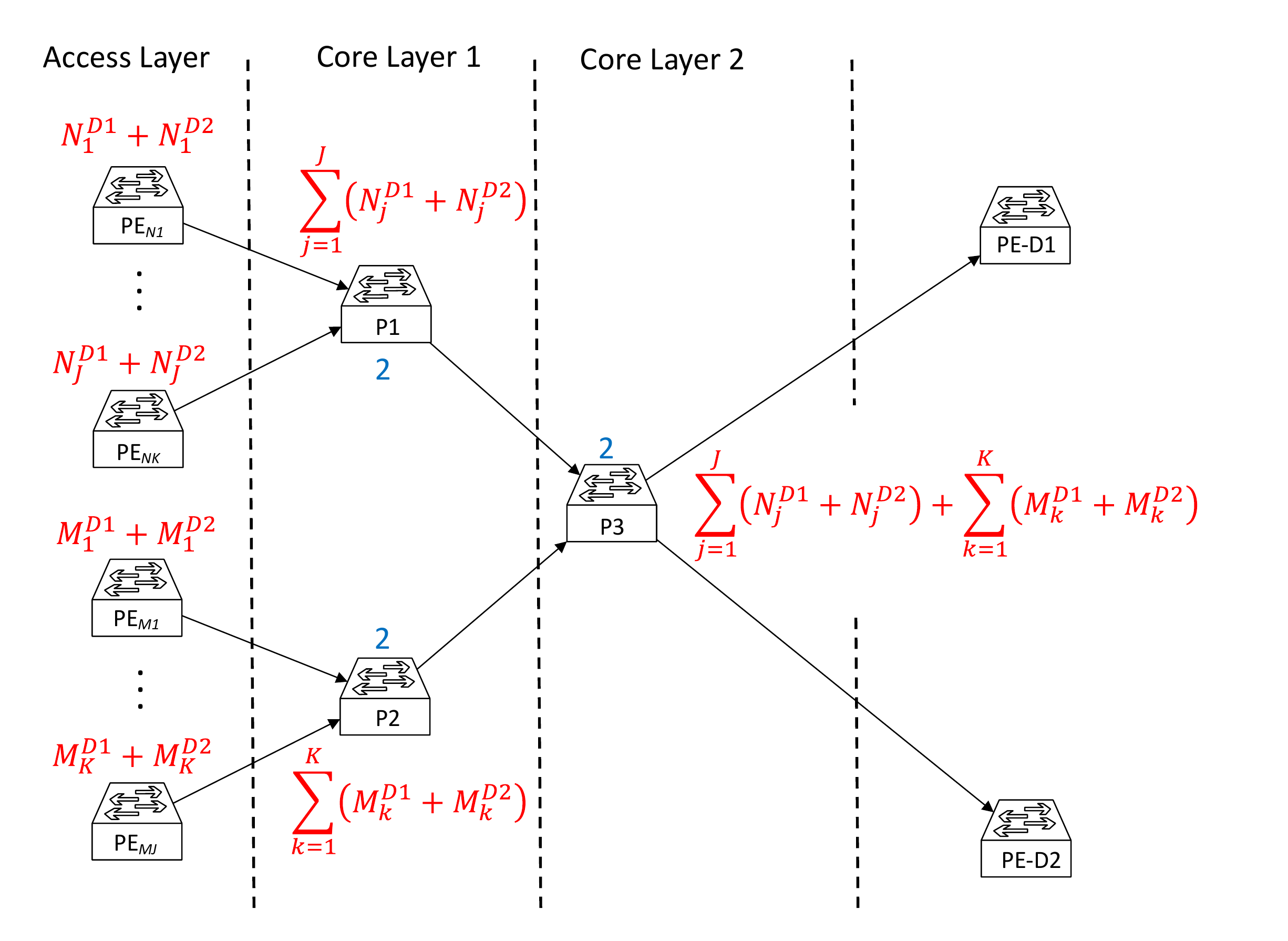}
    \vspace{-.5cm}
    \caption{Network structure and the related numbers of flows.} \label{fig:problemStatement}
\end{figure}

We distinguish the two types of flow-based switches performing flow forwarding, 
and divide network nodes into:
\begin{inparaenum}[(a)]
	\item \textit{Provider Edge} (PE) nodes; and
	\item \textit{Provider} (P) core nodes.
\end{inparaenum}
In Fig.~\ref{fig:problemStatement}, we present an example of a network topology where consecutive nodes perform traffic aggregation. Let us suppose that the whole traffic from ingress (domain entrance) nodes, i.e., PE$_{N1}$ to PE$_{NJ}$ and PE$_{M1}$ to PE$_{MK}$, is directed to egress (domain exit) nodes PE-D1 and PE-D2. The ingress nodes represent an access layer. A number of active flow entries in each PE is depicted in red, for example PE$_{N1}$ stores $N_1^{D1}+N_1^{D2}$ flow entries, where indices $D1$ and $D2$ represent which egress node the flows are directed to. At the first core layer, we observe a significant increase of the number of flows coming from the access layer. For instance, node P1 maintains as many as $\sum_{j=1}^J (N_j^{D1}+N_j^{D2})$ flows. In the second core layer, many more flow entries has to be served: $\sum_{j=1}^J (N_j^{D1}+N_j^{D2}) + \sum_{k=1}^K (M_k^{D1}+M_k^{D2})$. 

One of the main aims of the proposed mechanism is to reduce a number of flow entries in core switches (P nodes). Let us consider only flows directed to networks accessible via PE-D1. Since all flows from the access layer are directed to the single egress node (PE-D1), they can be represented by a single global label. If we consider two destinations, namely PE-D1 and PE-D2, we have to use two global labels. Due to our mechanism, ingress PE nodes are responsible for tagging each flow directed to a given destination egress PE node with a global unique label representing a particular egress node. In such the case, the number of flow entries is limited to exactly two per each P node in each core layer (depicted with blue). The number of flow entries in the access layer remains unchanged.

\subsection{Multipath Transmission}\label{subsec:multipath}

Since network traffic volume is continuously growing, one can expect that at some moment any network will experience congestions. To avoid this problem, one can use alternative paths leading to the same destination in the network. It is typical for mesh networks, that they contain more than one path between each source and destination. Such concurrent paths can be completely or partially disjoint. The use of concurrent paths can mitigate the problem of congested links. However, most of the legacy routing protocols do not implement multipath transmission. If they support multipath transmission, it is only based on Equal Cost MultiPath (ECMP)~\cite{rfc2992}. The notable exception amongst routing protocols is related to Cisco's EIGRP which can concurrently use paths with different costs~\cite{rfc7868}.

Our mechanism aims to use multipath transmission together with flow aggregation to avoid congestions in flow-based forwarding networks. Our proposal exploits many paths between same source-destination pair, but these paths do not need to be characterized with equal costs. Moreover, new paths are activated on demand only, that is, when a congestion appears and they do not tear down existing flows. Our mechanism searches the whole network to find the best new path avoiding congested links. The proposed solution extensively uses an aggregation procedure based on tagging with labels. 

\begin{figure}
    \centering
    \includegraphics[scale=0.7]{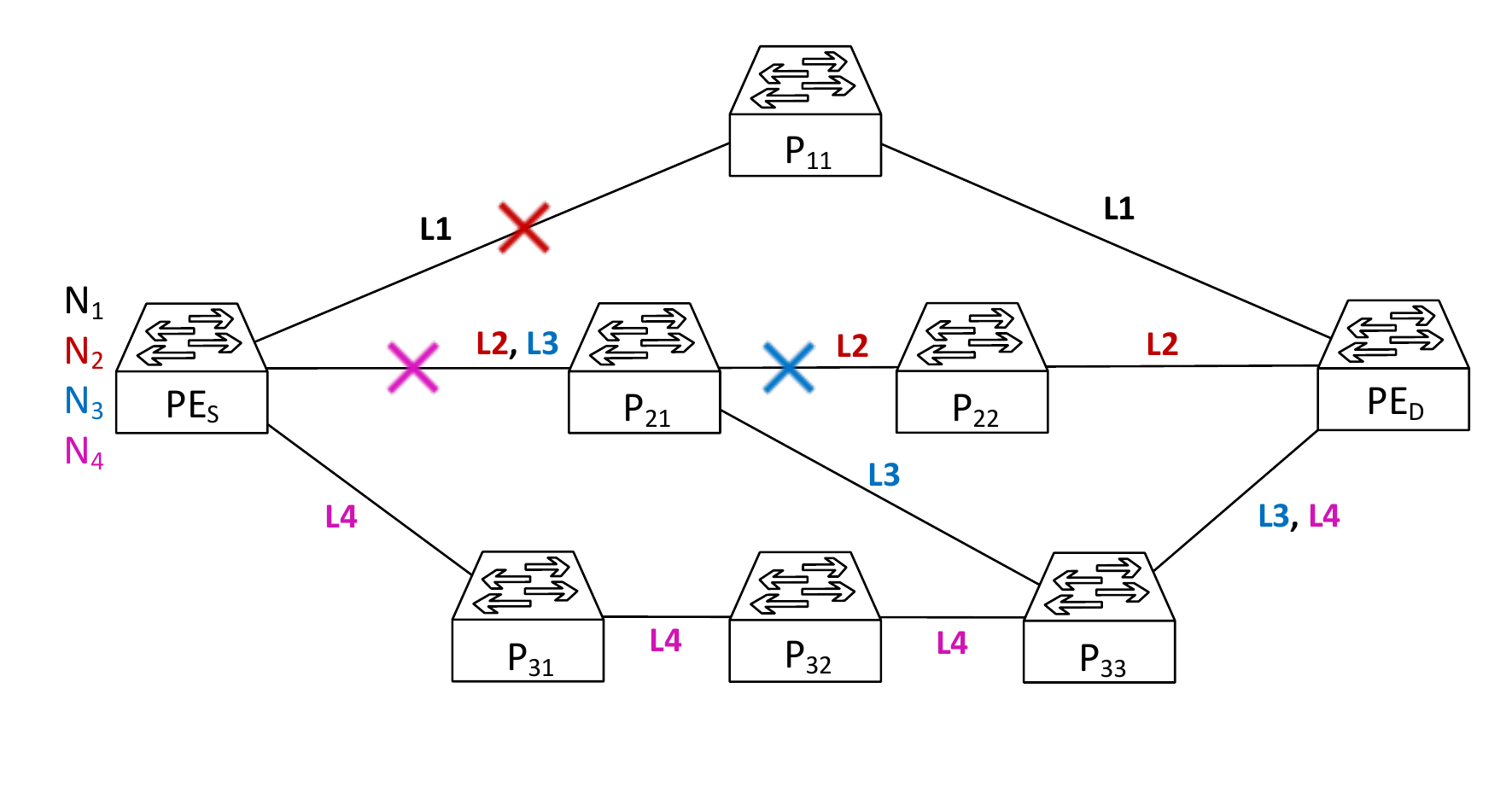}
    \vspace{-.5cm}
    \caption{Example used to explain the concept of multipath transmission adapted in the paper.} \label{fig:multipath}
\end{figure}

The concept of our multipath-based approach is here described with a simple exemplary network presented in Fig.~\ref{fig:multipath}. Sources of traffic are connected to the ingress node PE$_{\text{S}}$, while destination networks are attached to the egress node PE$_{\text{D}}$. One can observe that there exists a few paths connecting PE$_{\text{S}}$ and PE$_{\text{D}}$ nodes. Let us suppose that a routing procedure has chosen the path going through core node P$_{11}$. Being in line with our aggregation procedure, label L1 has been distributed enabling transfer along links marked with this label. All flows going from  PE$_{\text{S}}$ to PE$_{\text{D}}$ traverse this path and they are stamped  with MPLS label L1. All switches on the path forward packets according to this label. Now, in some moment a congestion appears on the link marked with a red cross. Our mechanism finds an alternative path (here: PE$_{\text{S}}$-P$_{21}$-P$_{22}$-PE$_{D}$) and distributes a new label L2 which will be used for packet forwarding. When all switches on this new path get this new label, ingress node PE$_{\text{S}}$ starts to mark new flows with label L2. The existing flows are still marked with label L1 and they still traverse the old path, namely PE$_{\text{S}}$-P$_{11}$-PE$_{D}$. The number of these flows is denoted by N$_1$ and does not increase. Label L2 is used by the number of flows denoted as N$_2$, and this number may increase since new flows arrive to the network. When next congestions appear in some links (in our example, the next congestion is indicated with a purple cross, and the subsequent one is indicated with a blue cross), new paths are found and new labels are distributed (L3 and L4, respectively). The similar scenario takes place after all congestions occur: existing flows use labels L1 and L2, while new flows use L3, and then --- label L4 (after the third congestion). The numbers of flows: N$_1$, N$_2$, N$_3$, related to the existing flows (using L1, L2, L3 labels, respectively) tend to decrease. Number N$_4$ of new flows may vary, but is likely to increase.

Thanks to the use of the flow-forwarding paradigm it is possible to distinguish existing flows from the new ones. Only a single active label is used for tagging new arriving flows directed to the same destination. The old labels are used for forwarding all flows existing before different congestions appeared. The tagging mechanism enables keeping existing flows on previously selected paths when the routing process chooses new paths between the given source-destination nodes. This mechanism also prevents increase of a number of flows from a particular source going via a congested link.

\subsection{Flow Installation}\label{subsec:insertion}

In SDN networks, two flow installation modes are available: reactive and proactive. In the former, each new flow reaching a switch generates signaling between the controller and an ingress switch resulting in on-demand rule installation, while in the latter forwarding rules are installed before flows arrive to the switch. A combination of these two modes is also possible.

The reactive flow insertion allows for flexible and fine-grained network management. However, such the approach involves a few serious limitations. First of all, every packet that does not have a match in a flow table of a switch has to be forwarded to an SDN controller. Then, the controller has to define actions for this packet and install a new rule for the next packets belonging to the particular flow in the switch. This situation may lead to overload of a controller with signalling messages \texttt{Packet\_IN}, especially in networks where a huge number of flows and switches is present, e.g., in DCs~\cite{benson, Kandula}. Also, other limitations have to be taken into account, therefore a limited number of requests per time unit can be handled by a single controller~\cite{nox,controller-performance}, thus decreasing network scalability. The reactive mode introduces an additional delay for every first packet of a new flow when it is forwarded to the controller. 
Moreover, it is likely that packets belonging to a single flow can arrive with such a high frequency that the installation of a forwarding rule in a switch takes place after many packets from the same flow arrive. This results in triggering many unnecessary \texttt{Packet\_IN} messages, causing further overload of the controller. Such the behavior can be exploited to attack an SDN controller.

On the other hand, the proactive flow insertion mode can easily mitigate these problems. It requires an advance knowledge about all traffic matches that could arrive into a switch. However, flexibility and precision of traffic control is lost in this case. Usually proactive rules are more general than those defined in reactive mode. This results from lack of knowledge about all traffic matches.

Existing flow-based switches suffer from the delays related to flow entry insertion into TCAM. This problem is mainly related to a weak management CPU and a slow communication channel present between the management of CPU and a switching chipset~\cite{commag-scalability, latency-sdn, measuring-latency}. These delays are especially cumbersome when network operates in the reactive mode~\cite{controller-performance}.

In the proposed mechanism, we limit signaling overhead, yet we still assume to apply a fine-grained flow forwarding. To install a new flow, we do not need to communicate with the SDN controller. This way, we exclude \texttt{Packet\_IN} messages. A controller proactively installs only rules for flow aggregates in a dedicated flow table in PE. Based on these patterns, a switch itself installs fine-grained flows without necessity to communicate with the controller. 
An SDN controller performs maintenance and modification only of aggregation rules. The mentioned modifications take place not often, i.e., only when a congestion appears and new paths are required. Introduction of these rules is feasible due to definition of a dedicated flow table. The detailed explanation of aggregation rules and switch behavior is given in Section~\ref{sec:mechanism}. 

%% file: sections/mechanism.tex
\section{Detailed Description of the Mechanism} \label{sec:mechanism}

Concerning the previously given partitioning of the switches, our mechanism assumes that:
\begin{itemize}
	\item \textit{Provider Edge} (PE) nodes map flows to MPLS labels; and
	\item \textit{Provider} (P) core nodes only forward packets according to MPLS labels.
\end{itemize}
We define a \textit{source-client network} (SCN) and a \textit{destination-client network} (DCN) as networks where sources and destinations of traffic are located, respectively. SCNs and DCNs are accessible via PE nodes only. 

To effectively map flows to the labels, the SDN controller builds and maintains a map of the physical topology and stores it in the form of the Link State Database (LSDB). LSDB is modified when a congestion starts to appear, i.e., link metrics are changed. The SDN controller calculates the best path only for pairs of PE nodes. The reverse Dijkstra algorithm (described in Section~\ref{sec:reverse-dijkstra}) is used to perform this task. For each PE, the controller allocates a global MPLS label representing a particular PE node. Then, the labels accompanied with information about proper output interfaces (obtained due to executing the shortest-path algorithm) are populated to each node. When a packet belonging to a particular flow reaches an ingress PE node, it is tagged with a proper MPLS label and subsequently it is forwarded to a pre-selected interface. This label indicates the egress PE node via which a particular DCN is reachable. Therefore, each node on the path will use a given label to reach the related particular PE node. Moreover, the same label will be used by any node in the whole network to reach the specified egress PE node. Such the approach results in flow aggregation and significant reduction of flow table entries in P nodes.

The proposed mechanism supports fast reaction to changes in traffic conditions. The SDN controller periodically collects utilization of links, and when any congestion in the network is recognized, the controller increases metrics of the overutilized links. Then, the reverse Dijkstra algorithm is recalculated for new flows that can appear between each pair of PE nodes. A new label for each PE is allocated. The controller populates all nodes with new labels and the related output interfaces. Therefore, only new flows use the new labels (i.e., the new paths). All the existing (old) flows are forwarded using the previously allocated labels (i.e., previously calculated paths). Such the approach stabilizes the flow forwarding process and introduces the multipath transmission. That is, between a selected source-destination PE pair the existing flows traverse the old paths and the newly recognized flows are redirected to the new paths.

\begin{figure}
    \centering
    \includegraphics[scale=0.8]{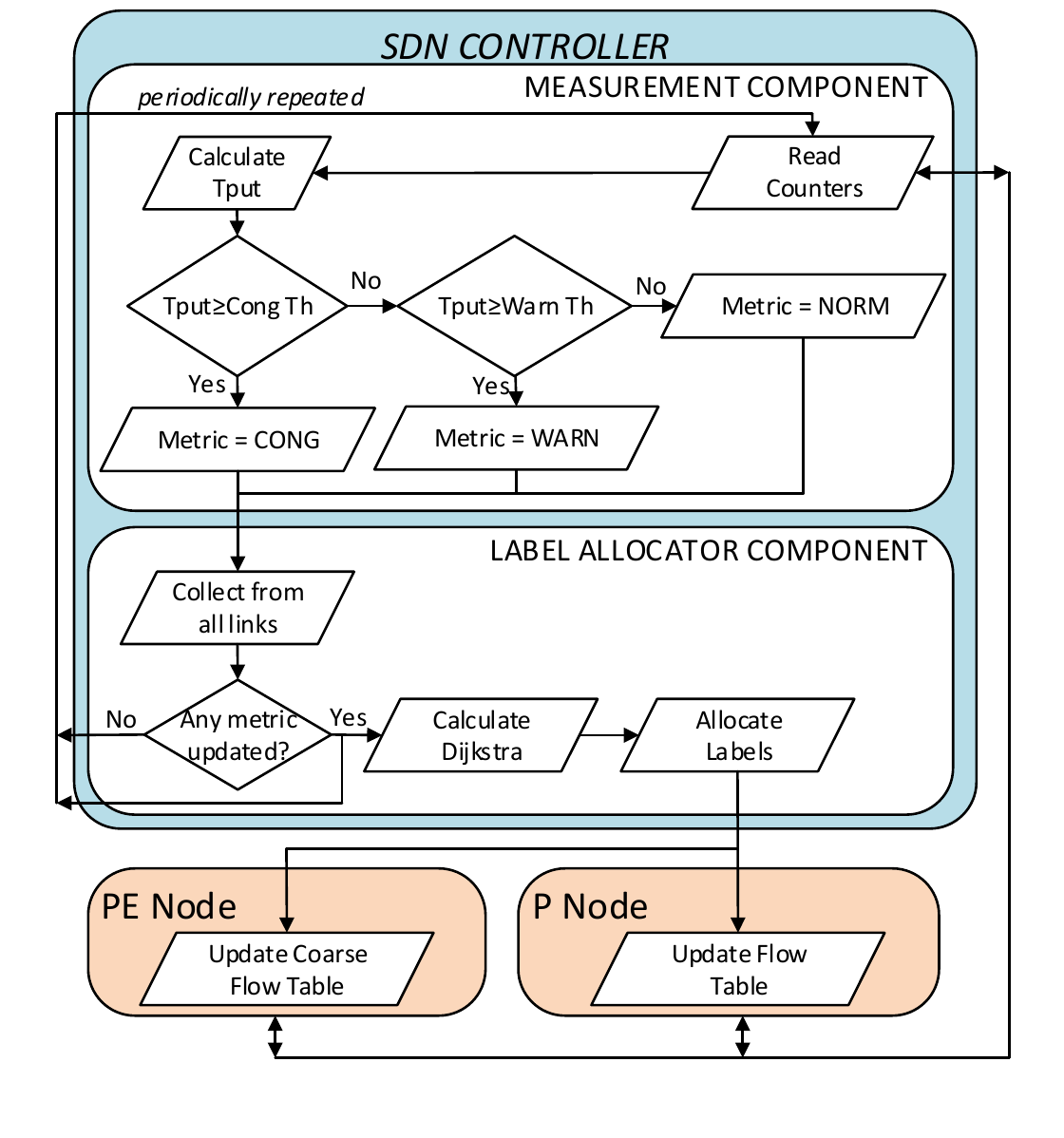}
    \vspace{-.5cm}
    \caption{The proposed mechanism.} \label{fig:algoritm}
\end{figure}

The proposed management system running at the SDN controller (see~Fig.~\ref{fig:algoritm}) is logically divided into the two components:
\begin{itemize}
	\item \textit{Measurement Component} responsible for gathering link utilization and modification of metrics; and
	\item \textit{Label Allocator Component} calculating paths and distributing MPLS labels.
\end{itemize}
Below, first we describe the way the packets and flows are processed in various types of network nodes, and then we describe the operation of the both defined components responsible for defining how the nodes should process the data.

\subsection{Flow Processing in PE and P Nodes} \label{sec:flow-processing}

Each PE node implements flow-based forwarding.
The way the flows are defined is neutral from the viewpoint of our mechanism. For instance, a traditional 5-tuple (source and destination adresses/ports with the L4 protocol) can be used.

\begin{figure}
    \centering
    \includegraphics[scale=0.8]{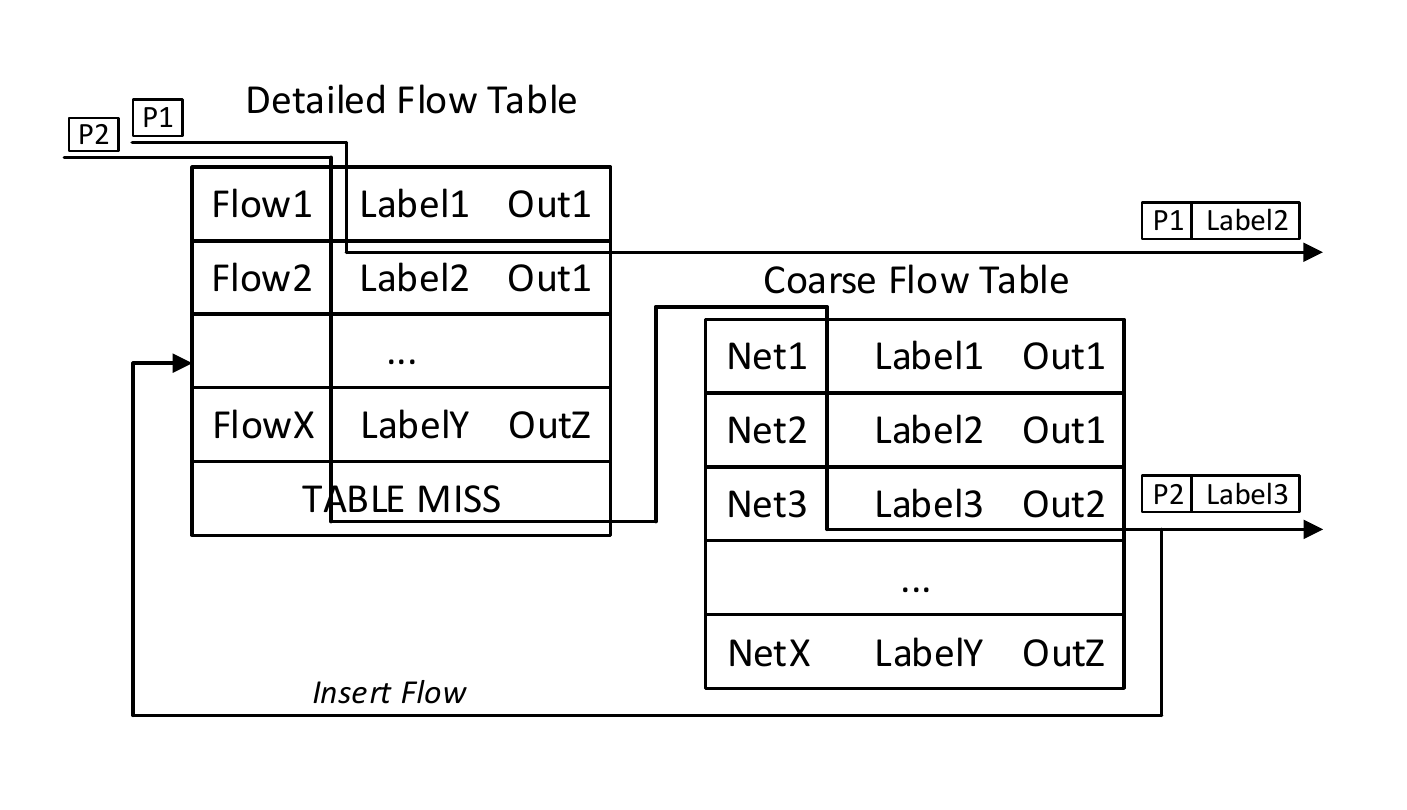}
    \caption{Packet processing pipeline in an edge (PE) node.}\label{fig:packet}
\end{figure}

In accordance with the OpenFlow specification, we propose to use the two
flow tables in each PE node. The \emph{Detailed Flow Table} (DFT) stores the detailed information on active flows. The \textit{Coarse Flow Table} (CFT) contains the mapping between DCNs and pairs (\texttt{output MPLS label}, \texttt{output interface}). Thus, when a packet reaches a PE node, it is processed following the pipeline shown in Fig.~\ref{fig:packet}.

We consider the following two cases.
\begin{enumerate}
    \item If a packet (P1 in the figure) matches an existing flow in the DFT, then this packet is processed according to the list of actions present for such an entry. That is, the packet is tagged with a selected MPLS label and forwarded to the indicated output interface.
    \item If a packet (P2 in the figure) does not match DFT, then this packet is redirected to the CFT. It contains entries composed of a DCN and a list of actions (i.e., push a pre-defined MPLS label and direct the packet to a pre-selected output interface). Thus, when the match is found, the specified actions on the packet are performed 
	and the detailed flow entry is created in the DFT. The entry is based on information gathered from the packet's header fields. Hence, for the new flow defined on the basis of this header, the entry action list is copied from the CFT. Idle timeout of this entry is set to a finite pre-defined value. The issue of timeouts setting and usage is explained in detail in Subsection~\ref{sec:garbage}.
\end{enumerate}

In legacy OpenFlow protocol only an SDN controller may insert flow entries into flow tables. As mentioned before, such the procedure may lead to a storm of \texttt{Packet\_IN} messages received by a controller. When we consider huge networks carrying millions of flows, the reactive flow insertion may lead to overload of the controller. Therefore, in our mechanism we improve the standard operations of the OpenFlow protocol, and in result we reduce the number of messages exchanged between the controller and switches. The CFT contains general rules indicating how flows should be processed. These rules in some way are persistent and are updated by a controller, only when a congestion appears. The number of CFT entries is related to the number of DCNs. Insertion of a particular granular flow entry into DFT is made by PE switch on its own (as presented in Fig.~\ref{fig:packet}). The proposed improvement removes the need to use \texttt{Packet\_IN} messages, but still enables reactive treatment of flows.

For P nodes, only a single flow table is required. When a packet reaches such a node, this packet is matched on the basis of a label only.
Then, the packet is sent out to a proper output interface with exactly the same label.
If a legacy MPLS router is used as a P node, such a node performs only the ordinary label swapping operation (the output label is swapped to the input label).

\subsection{Measurement and Label Allocator Components}

The Measurement Component (MC) periodically retrieves data from link counters.
The collected data is used to calculate the bandwidth utilization at links. There are two utilization thresholds configured: warning (\texttt{WarnTh}) and congestion (\texttt{CongTh}). The latter is determined with a value larger than the former. Each time the amount of the traffic throughput (\texttt{Tput}) exceeds one of the defined thresholds, the MC changes a link metric in the LSDB stored by the Label Allocator Component (LAC). 
We propose the three following values of the configurable link metrics related to the threshold values.
\begin{inparaenum}[(1)]
	\item \texttt{NORM}: Normal Metric (a default IGP metric) for the link utilization with value not greater than the \texttt{WarnTh} threshold;
	\item \texttt{WARN}: Warning Metric for the link utilization with value between the \texttt{WarnTh} and \texttt{CongTh} threshold;
	\item \texttt{CONG}: Congestion Metric for the link utilization exceeding the \texttt{CongTh} threshold.
\end{inparaenum}

LAC builds and maintains LSDB.
Each time the MC changes any link metric, the recalculation of the shortest paths is triggered for each of the PE nodes treated as a root. The number of such recalculations can be limited to some PE nodes only. We use reverse Dijkstra algorithm based on the Dijkstra algorithm. Below, in Sec.~\ref{sec:reverse-dijkstra} we describe this process in detail. After recalculation, a new label set is allocated. The maximum size of this set equals the number of PE nodes. After consecutive recalculations of the reverse Dijkstra algorithm, only one unique label represents a destination PE for new flows. The existing flows are still forwarded with the previously allocated labels.

As presented in Sec.~\ref{sec:flow-processing}, for PE nodes the CFT table is updated and filled in by the SDN controller. The CFT contains entries composed of an address of a DCN and a list of actions (i.e., push MPLS label and send it to an output interface) with infinite timeout. Each time recalculation of the reverse Dijkstra algorithm is triggered, it results in the CFT update. The old list of actions for each DCN is replaced with a new label and a new output interface (based on the structure of the new shortest-path tree).

After the recalculation, new label entries in flow tables are proactively installed in the P nodes as well. A single entry of this kind contains a match based only on a new input MPLS label and the forwarding action: the output interface is based on the currently calculated shortest-path tree. The idle timeout is set to infinity.

\subsection{Possible Extensions}\label{sec:extension}
The proposed mechanism does not limit functionalities which are present in OpenFlow. All OpenFlow actions still can be performed. The only aspect which differs our solution from the standard OpenFlow behavior consists in adding the \texttt{Insert Flow} action. This action is taken by a switch itself and results in a flow insertion into DFT on the basis of an entry transferred from CFT.

\begin{figure}
    \centering
    \includegraphics[scale=0.75]{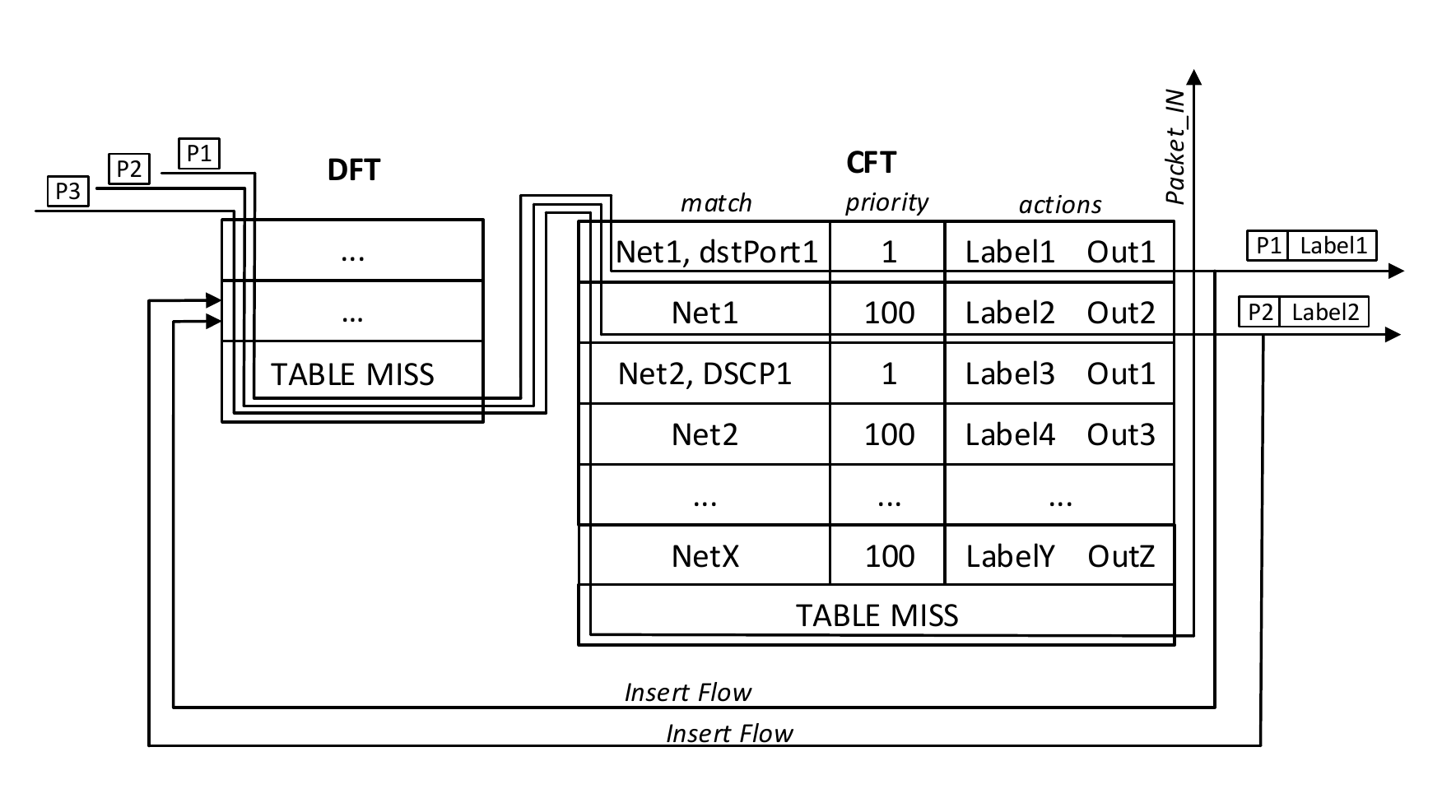}
    \caption{Extension of Coarse Flow Table.}\label{fig:application}
\end{figure}

The match rule present in CFT does not need to be based on a destination network only. It can be composed of any combination of fields and wildcards supported by OpenFlow. In Fig.~\ref{fig:application} we depicted a few exemplary match rules. Let us consider three packets arriving to the switch, i.e., P1, P2, P3. All of them does not match any entry in DFT, thus they are redirected to CFT. P1 and P2 are destined to the same network, but P1 also matches an extended rule with a destination Layer 4 port, hence it is send out to a different output port with a different label (\texttt{Label1}, \texttt{Out1}) than in the case of P2 (\texttt{Label2}, \texttt{Out2}). Such the approach allows to serve distinct applications in a specific way. Another possibility of packet serving is a usage of the DSCP field to fulfill QoS requirements. It is also possible to control traffic directly by an SDN controller. If there is a particular type of traffic that is expected to be managed by an SDN controller, CFT should posses a Table Miss entry. This entry allows to generate a \texttt{Packet\_IN} message. After packet is analysed, a controller installs an appropriate entry in DFT.

\subsection{Reverse Dijkstra: the Recalculation Algorithm} \label{sec:reverse-dijkstra}

Whenever a congestion appears, calculation of new paths avoiding the overloaded links is needed. For finding these new paths, a form of the Dijkstra algorithm is used. In the proposed mechanism, we do not need to perform path recalculation for all network nodes. To decide for which PE nodes this recalculation is necessary, the mechanism starts the investigation of labels used by the packets transferred via the link at which a new overload has just been recognized. Each of these labels indicates a specific destination PE node, called thereupon an `affected PE' node. To avoid the negative impact of the experienced congestion, new paths directed to the affected PEs should be found. Consequently, the new labels have to be allocated. For the non-affected PE nodes, path recalculation and label reallocation are not needed.

\begin{figure}	
	\centering
    \subfloat[Exemplary network topology.~~~~\label{fig:reverseDijkstra}]{\includegraphics[scale=0.7]{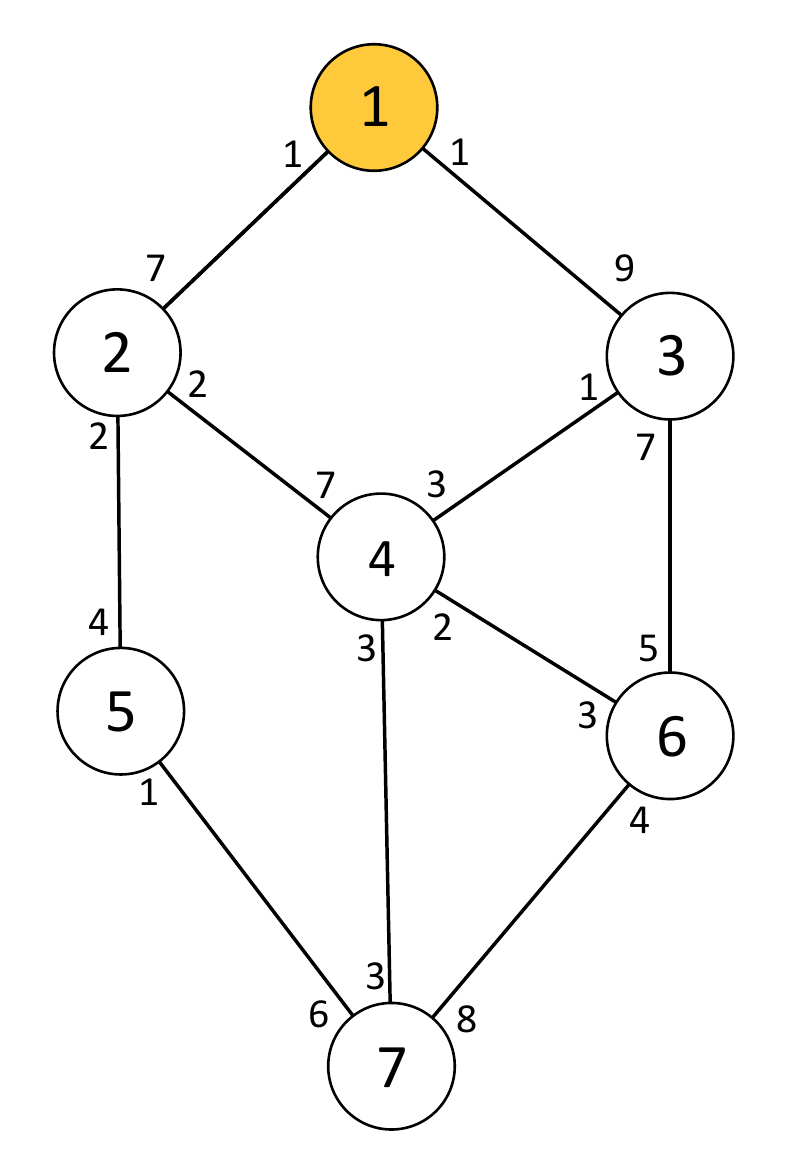}}\hspace{0.2cm}
    \subfloat[Reverse Dijkstra tree.\label{fig:reverseDijkstraTree}]{\includegraphics[scale=0.7]{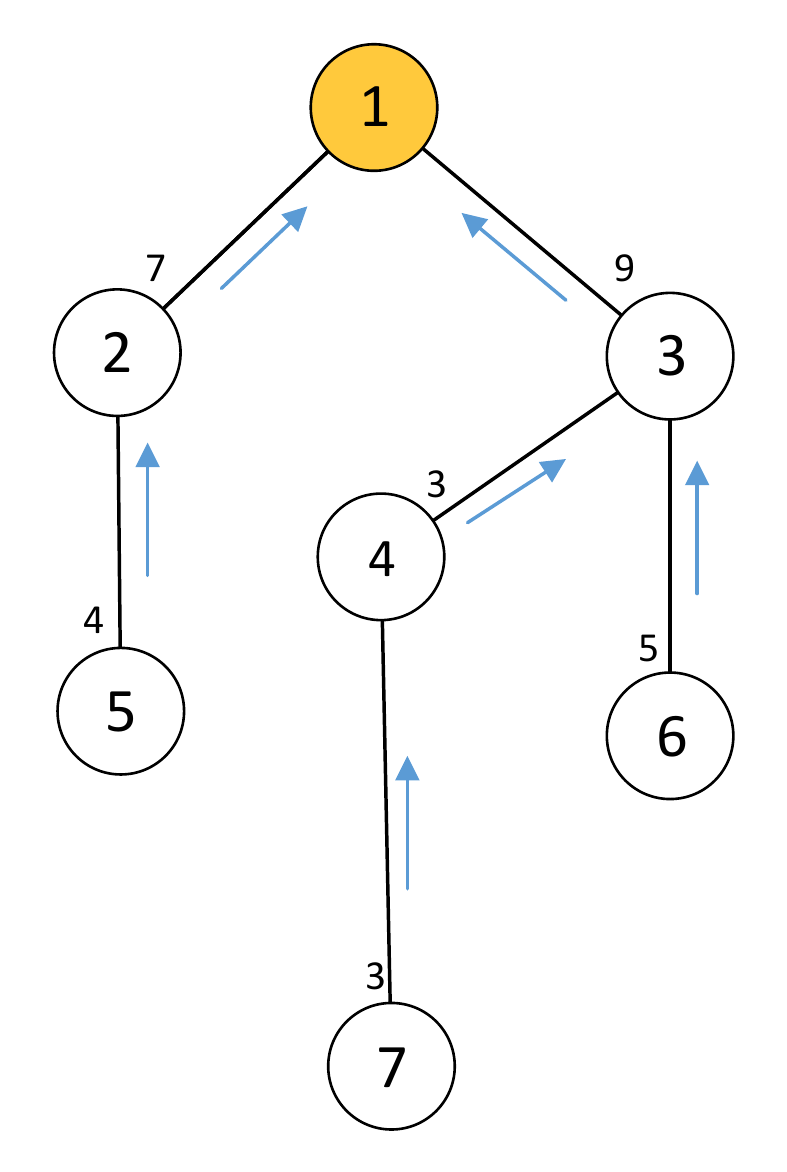}}
    \caption{An example for the reverse Dijkstra procedure.}\label{fig:reverseExample}
\end{figure}

In the presented mechanism, we calculate paths from the perspective of each affected PE node treated as a root. However, the weights used in the shortest-path algoritm are related to the links directed in the opposite way (i.e., towards the root). Therefore, we call this procedure `reverse Dijkstra'. 
For better explanation how this procedure works, in Fig.~\ref{fig:reverseExample} we present an examplary network topology together with the obtained reverse Dijkstra tree. The destination PE (a root for the reverse Dijkstra calculation) is colored with orange (node 1). A metric of each link for each direction is depicted in Fig.~\ref{fig:reverseDijkstra}. For example, if we consider connection between nodes 1 and 2, a regular Dijkstra algorithm uses metric 1, while the reverse Dijkstra will use metric 7. In Fig.~\ref{fig:reverseDijkstraTree} the outcome of the whole reverse Dijkstra procedure is presented in the form of the tree with the used metrics. With blue arrows we indicated the traffic direction. MPLS label directed to the destination node 1 will be distributed down the reverse Dijkstra tree. In case of a regular Dijkstra algorithm one has to perform six Dijstra calculations using nodes 2-7 as roots.

\subsection{Flow Garbage Procedure} \label{sec:garbage}

Flow tables require maintenance to remove unused entries.
We propose to apply a standard OpenFlow procedure for flow entry removal from the DFTs in PE nodes. That is, an idle timeout counter is used for this purpose. 
Some finite value is assigned for each flow present in DFT, while rules placed in CFT always have infinite timeout.

For the P nodes we propose a procedure aligned with the current functionality of OpenFlow. Namely, when an SDN controller calculates new paths and allocates a new label to a particular egress PE, the related entries are proactively installed with infinite idle timeout into forwarding tables of P switches belonging to these paths. Simultaneously, in P switches, the controller modifies previous rules destined to this PE (identified by previously allocated labels), i.e., the controller changes only timeout timers from infinity to a finite value. Thus, all existing flows are forwarded without changes. When old flows end, their idle timeout counters will exceed and the removal of such flows from the flow tables will take place. When all the flows related to a particular label expire in all P nodes, this label returns to the pool of available labels used by the SDN controller. The infinite value set to the  idle timeout of flow entries in P switches is needed to sustain readiness to handle flows, even after a long absence of any related traffic.

\subsection{Integration with Existing Off-the-Shelf Technologies}

A full upgrade of network devices may result in huge capital expenditures for an operator. An incremental approach can span this task in time. In this subsection, we summarize the ideas how to integrate our mechanism with existing off-the-shelf technologies.

In the proposed system, we distinguish two types of network nodes: PE and P. The former has to be upgraded, while the latter may be a legacy MPLS router. The PE nodes have to operate with flow-based forwarding, thus they have to work in accordance with the proposed functionalities (i.e., OpenFlow switches will conform to the procedure of a flow insertion between tables, as shown in Fig~\ref{fig:packet}).

Since in our mechanism P nodes forward traffic according to MPLS labels, a network operator does not need to replace legacy routers if they support MPLS. The labels used in our mechanism have the global meaning. The application of centralized management offered by SDN controllers enables synchronization of label distribution to all the P and PE nodes. An SDN controller distributes a unique global label related to a particular egress PE node, and then a simple label swapping is applied. The input label has to be swapped to the same label. Each time a controller recalculates paths and allocates new labels, it reconfigures static Label Switched Path (LSP), hence a static label binding can be used. In such the case an SDN controller may configure routers using for instance NetConf, SNMP, XMPP or Segment Routing.
Standard signalling mechanisms, such as LDP, RSVP, cannot be applied because due to these mechanisms MPLS labels are assigned by each node independently of others, and in consequence the labels have a local meaning only.

\begin{table}[h]
\centering
\caption{Signaling protocols between an SDN controller and network nodes.}
\label{tab:protocols}
\begin{tabular}{|c|c|c|c|}
\hline
\multicolumn{2}{|c|}{Counter readouts} & \multirow{2}{*}{\begin{tabular}[c]{@{}c@{}}Flow\\management\end{tabular}} & \multirow{2}{*}{\begin{tabular}[c]{@{}c@{}}Topology \\ discovery\end{tabular}} \\ \cline{1-2}
Push              & Pull               &                                                                              &                                                                                \\ \hline \hline
NetFlow           & OpenFlow           & OpenFlow                                                                     & LLDP                                                                           \\ \hline
IPFIX             & SNMP               & SNMP                                                                         & OSPF                                                                           \\ \hline
sFlow             &                    & NetConf                                                                      & IS-IS                                                                          \\ \hline
jFlow             &                    & XMPP                                                                         &                                                                                \\ \hline
                  &                    & Segment Routing                                                              &
\\ \hline
\end{tabular}
\end{table}

Since our solution represents a measurement-driven mechanism, it needs to collect some link statistics. They can be gathered and communicated with use of various protocols depending on functionality supported by switches and routers, and the assumed method of obtaining counter readouts (i.e., push or pull). For the push method, protocols such as NetFlow, IPFIX, sFlow, jFlow may be used. These protocols are designed to periodically report traffic statistics. For the pull method, one can apply OpenFlow or SNMP. These protocols offer on-demand acquisition of statistics.

As controllers have to maintain LSDBs, they need to discover the network topology. Information collected from the well-known protocols such as LLDP, OSPF, IS-IS can be used to build LSDB in an SDN controller. In Tab.~\ref{tab:protocols} we summarize some market-available protocols that can be applied with our mechanism.

%% file: sections/evaluation.tex
\section{Evaluation} \label{sec:evaluation}

This section presents simulation setups and results for performance evaluation of the proposed mechanism. All the tests were run on ns-3 simulator~\cite{ns3}. To conduct the evaluation, we implemented components described in Sec.~\ref{sec:mechanism}. 
A new MPLS module offering
concurrent processing of IP packets and MPLS frames was implemented as well. Moreover, we have added features enabling: SDN-based central management of a network, LSDB maintenance, the reverse Dijkstra algorithm calculation, the new MPLS label distribution procedure, and functionalities related to the Measurement Component. 

\begin{figure}
     \centering
     \includegraphics[scale=0.3]{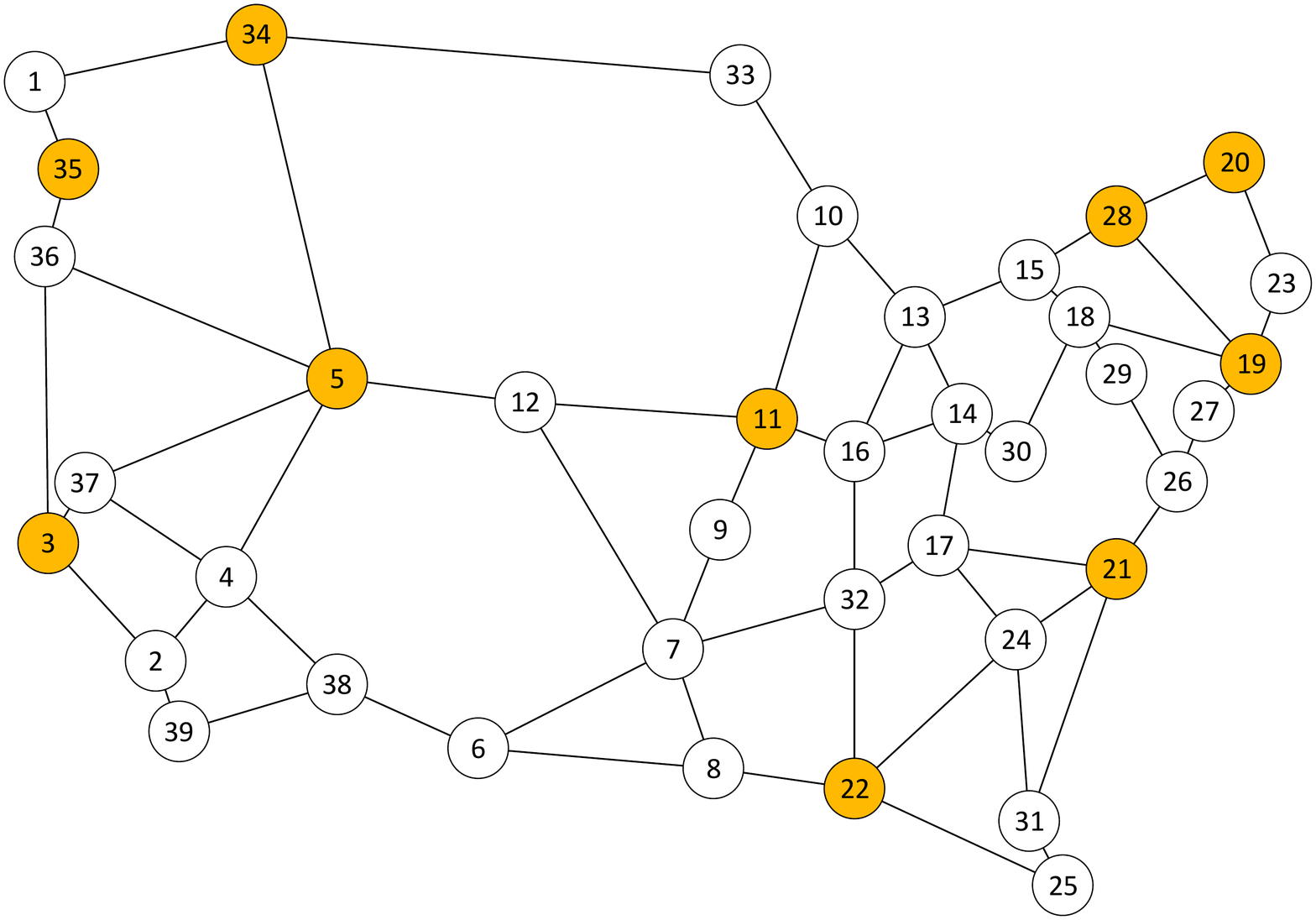}
     \caption{US backbone network used in the numerical evaluation. The yellow vertices represent SCNs/DCNs.}\label{fig:topo}
 \end{figure}

We used the US backbone network~\cite{us-net} for all the experiments. The network contains 39 nodes and 61 bidirectional links. 10 selected nodes (PEs) serve as attachment points for traffic sources and destinations playing the both SCN and DCN roles simultaneously (as depicted in Fig.~\ref{fig:topo}). All links connecting network nodes are set to 100\,Mbps with 1\,ms transmission delay. Links interconnecting SCNs/DCNs with PE are set to 1\,Gbps with 1\,ms transmission delay. Such the configuration allows to avoid bottlenecks in the access part of the network. %

We study transmission of the TCP traffic only. Network traffic is injected with use of the ns-3 internal random number generators. Flow sizes are generated on the basis of the Pareto distribution with the shape parameter equal to 1.5 and the mean value set to 500\,kB. Flow inter-arrival times were selected in accordance with exponential distribution (the mean value equals 3\,ms). The simulation time was set to 100\,sec. Data collection was started after elapsing of the first 10\,sec. of 
the simulation warm-up time.

Simulations were conducted for a few combination of the (\texttt{CongTh}, \texttt{WarnTh}) pairs: $(0.8,0.4-0.7)$, $(0.85,0.4-0.7)$, $(0.9,0.4-0.7)$, where the warning threshold was increased with $0.1$ step. Each simulation was repeated 20 times. Then, the 95\% confidence intervals were computed. For each pair of the thresholds, we use the same set of seed values to achieve repeatable traffic conditions. Additionally, such the procedure enables us to carry a fair comparison among different setups. For all the simulation setups, we fixed the following values of link metrics placed in LSDB: \texttt{NORM} = 1, \texttt{WARN} = 1000, \texttt{CONG} = 65535.

Moreover, to observe the gain, we performed 20 simulations with the same set of seeds and the proposed mechanism disabled. This scenario is known as `Legacy SDN', and by it we mean operation of standard OpenFlow switches, all working in the reactive mode (each switch on the path always sends the \texttt{Packet\_IN} message to the controller when a new flow arrives). For this case, we take an assumption that the controller allocates flows according to a regular Dijkstra tree, that is multipath transmission is not used.

Tab.~\ref{tab:results} presents the data obtained for evaluation of our mechanism. This data was collected from all the nodes. On this basis, we are able to calculate:
\begin{inparaenum}[(a)]
	\item the total number of the transmitted (`Tx') and received (`Rx') bytes;
	\item the percentage of dropped packets (`Drop Pkts'),
	\item the mean achieved network throughput (`Avg Tput').
\end{inparaenum}
Moreover, in the table we include a parameter expressing the received data gain (`$\mathit{Rx}_{\text{Gain}}$') obtained in relation to the situation without use of the proposed mechanism, as shown in the following equation:
\begin{equation}
	\mathit{Rx}_{\text{Gain}} = \frac{\mathit{Rx} - \mathit{Rx}_{\text{LegacySDN}}}{\mathit{Rx}_{\text{LegacySDN}}} \times 100\%,
\end{equation}
where $\mathit{Rx}$ means the total received data during simulation when the mechanism is used, and $\mathit{Rx}_{\text{LegacySDN}}$ expresses the total received data without application of the mechanism.

To estimate the scalability of our mechanism, we gather the total number of flow entries on all the PE nodes per second (`Sum of DFT entries (PE)'), and the mean number of labels present on a single P node per second (`Avg label entries (P)'). To show the efficiency of flow processing supported by our tagging approach, during the whole simulation we observe the number of label entries present on all the P switches and we store the maximum values. In the table, the mean of maximum values over all simulations is provided (`Max label entries (P)'). Furthermore, for the evaluation of flow processing scalability of the proposed mechanism, we define the following indicator: the maximum Flow Reduction Indicator (`$\mathit{maxFRI}$').
Its calculation is performed as follows. First, we distinguish flow tuples and labels. The flow tuples are used by PE switches, while labels represent aggregated flows. The latter are used by the P switches for forwarding. Supposing that all the PE nodes serve as traffic sources, in each step of the simulation we observe the total number of flows in the network. Second, we verify how many labels on a single P node are active due to presence of the mentioned flows. Third, we calculate the average number of active labels per a single P node. This value shows the mean number of labels used for traffic processing in the core of a network. Finally, to present a single indicator for the whole simulation time ($simTime$), we use the maximum value to calculate $\mathit{maxFRI}$ as shown below:

\begin{gather}
  	\mathit{maxFRI} = \max\limits_{\mathit{simTime}}\left[1-\frac{\avg(\#\mathit{labels}_{_P})}{\sum\limits_{\mathit{PE}}{\#\mathit{flows}_{_{\mathit{PE}}}}}\right] \times 100\%. \label{eq:max-fri} 
\end{gather}
This value exemplifies the maximum percentage reduction of the number of entries used by switches in flow tables in comparison to legacy flow switching (without any aggregation procedure). 
This number expresses the decrease rate of flow table entries when our mechanism is used. We want to stress that the $\mathit{maxFRI}$ represents the situation when all flows from all the PE nodes are present on all core P switches. This is a Key Performance Indicator (KPI) which enables us to quantify efficiency of flow reduction during the system operation. This KPI is based on measurements performed during the system run. No other mechanisms are compared with use of this indicator.

We also define another indicator, known as Comparative Flow Reduction Indicator (`$\mathit{CFRI}$'), which measures efficiency of flow reduction for core nodes when simulated scenarios are compared. Contrary to $\mathit{maxFRI}$,
$\mathit{CFRI}$ uses simulation data gathered in the both compared mechanisms. The indicator is defined as follows:
\begin{gather}
  	\mathit{CFRI} = \left[1-\frac{\avg(\#\mathit{labels}_{_P})}{\avg(\#\mathit{flows}_{\text{LegacySDN}_P})}\right] \times 100\% \label{eq:fri},
\end{gather}
where $\avg(\#\mathit{labels}_{_P})$ is the average number of flow entries (per second) in a core P node when our mechanism is applied, and $\avg(\#\mathit{flows}_{\text{LegacySDN}_P})$ represents the average number of flow entries (per second) in a core P node when the `Legacy SDN' scenario is considered.

\begin{center}
\begin{sidewaystable}
  \centering
  \caption{Evaluation results}
  \resizebox{1.0\columnwidth}{!}{%
    \begin{tabular}{c|c||c|c|c|c||c|c|c|c|c}
    \toprule
    \multirow{2}[2]{*}{\texttt{CongTh}} & \multirow{2}[2]{*}{\texttt{WarnTh}} & Tx    & Rx    & Drop & Avg Tput  & $\mathit{Rx}_{\text{Gain}}$ & Sum of DFT & Avg label & Max label & $\mathit{maxFRI}$ \bigstrut[t]\\
          &       & [GB]  & [GB]  &  Pkts [\%] & [Mbps] & [\%] & entries (PE) & entries (P) & entries (P) & [\%] \bigstrut[b]\\
    \midrule
    \multirow{4}[8]{*}{0.8} & 0.4   & 15.3$\pm$0.10 & 14.5$\pm$0.10 & 5.4$\pm$0.05 & 1315.9$\pm$8.9 			& 32.2$\pm$0.7 & 13377.6$\pm$13.8 	& 81.1$\pm$0.9 & 92.6$\pm$3.6 & \textbf{99.55$\pm$0.01} 			\bigstrut\\
\cmidrule{2-11}          & 0.5   	& 15.3$\pm$0.09 & 14.4$\pm$0.09 & 5.6$\pm$0.04 & 1312.8$\pm$8.5 			& 31.9$\pm$0.6 & 13741.1$\pm$11.1 	& 84.9$\pm$0.6 & 96.5$\pm$2.7 & \textbf{99.55$\pm$0.01} 	\bigstrut\\
\cmidrule{2-11}          & 0.6   	& 15.2$\pm$0.10 & 14.4$\pm$0.09 & 5.6$\pm$0.04 & 1308.2$\pm$8.5 			& 31.4$\pm$0.6 & 13951.9$\pm$9.2 		& 87.5$\pm$0.7 & 99.8$\pm$3.9 & 99.54$\pm$0.01 		\bigstrut\\
\cmidrule{2-11}          & 0.7   	& 15.2$\pm$0.10 & 14.4$\pm$0.09 & 5.6$\pm$0.04 & 1306.1$\pm$8.5 			& 31.2$\pm$0.6 & 13922.1$\pm$12.6 	& 85.8$\pm$0.6 & 99.1$\pm$1.6 & \textbf{99.55$\pm$0.01} 			\bigstrut\\
    \midrule			
    \multirow{4}[10]{*}{0.85} & 0.4 & 15.3$\pm$0.11 & 14.5$\pm$0.10 & 5.3$\pm$0.04 & 1318.7$\pm$9.4 			& 32.5$\pm$0.7 & 13139.1$\pm$10.3 	& 80.0$\pm$0.5 & 91.0$\pm$3.2 & \textbf{99.55$\pm$0.01} 			\bigstrut\\
\cmidrule{2-11}          & 0.5   	& 15.3$\pm$0.10 & 14.5$\pm$0.09 & 5.5$\pm$0.03 & 1315.7$\pm$8.5 			& 32.2$\pm$0.6 & 13497.9$\pm$10.3 	& 84.7$\pm$0.6 & 96.7$\pm$3.4 & 99.53$\pm$0.01 			\bigstrut\\
\cmidrule{2-11}          & 0.6   	& 15.2$\pm$0.09 & 14.4$\pm$0.09 & 5.6$\pm$0.04 & 1310.4$\pm$7.8 			& 31.7$\pm$0.7 & 13776.9$\pm$10.7 	& 87.5$\pm$0.6 & 100.1$\pm$2.2 & 99.52$\pm$0.01 			\bigstrut\\
\cmidrule{2-11}          & 0.7   	& 15.2$\pm$0.09 & 14.4$\pm$0.09 & 5.6$\pm$0.04 & 1306.4$\pm$8.1 			& 31.3$\pm$0.7 & 13837.7$\pm$10.5 	& 88.1$\pm$0.8 & 101.5$\pm$4.0 & 99.52$\pm$0.01 					\bigstrut\\
    \midrule
    \multirow{4}[10]{*}{0.9} & 0.4  & 15.3$\pm$0.09 & 14.5$\pm$0.09 & 5.4$\pm$0.04 & \textbf{1319.8$\pm$7.9} 	& 32.6$\pm$0.7 & 13133.6$\pm$14.5 	& 81.5$\pm$0.8 & 93.0$\pm$2.9 & 99.54$\pm$0.01 					\bigstrut\\
\cmidrule{2-11}          & 0.5   	& 15.3$\pm$0.10 & 14.5$\pm$0.10 & 5.5$\pm$0.03 & 1316.8$\pm$9.0 			& 32.3$\pm$0.6 & 13392.9$\pm$10.4 	& 85.0$\pm$0.6 & 97.3$\pm$2.8 & 99.53$\pm$0.01 			\bigstrut\\
\cmidrule{2-11}          & 0.6   	& 15.3$\pm$0.10 & 14.4$\pm$0.10 & 5.6$\pm$0.05 & 1313.2$\pm$8.7 			& 31.9$\pm$0.6 & 13617.3$\pm$13.8 	& 88.1$\pm$0.6 & 100.7$\pm$3.7 & 99.52$\pm$0.01 					\bigstrut\\
\cmidrule{2-11}          & 0.7   	& 15.3$\pm$0.10 & 14.4$\pm$0.09 & 5.6$\pm$0.05 & 1311.0$\pm$8.4 			& 31.7$\pm$0.8 & 13673.4$\pm$14.5 	& 89.0$\pm$0.8 & 102.5$\pm$3.1 & 99.51$\pm$0.01 					\bigstrut\\
    \bottomrule
    \end{tabular}%
    }
  \label{tab:results}%
\end{sidewaystable}
\end{center}

\begin{figure}
    \centering
    \includegraphics[width=0.7\columnwidth]{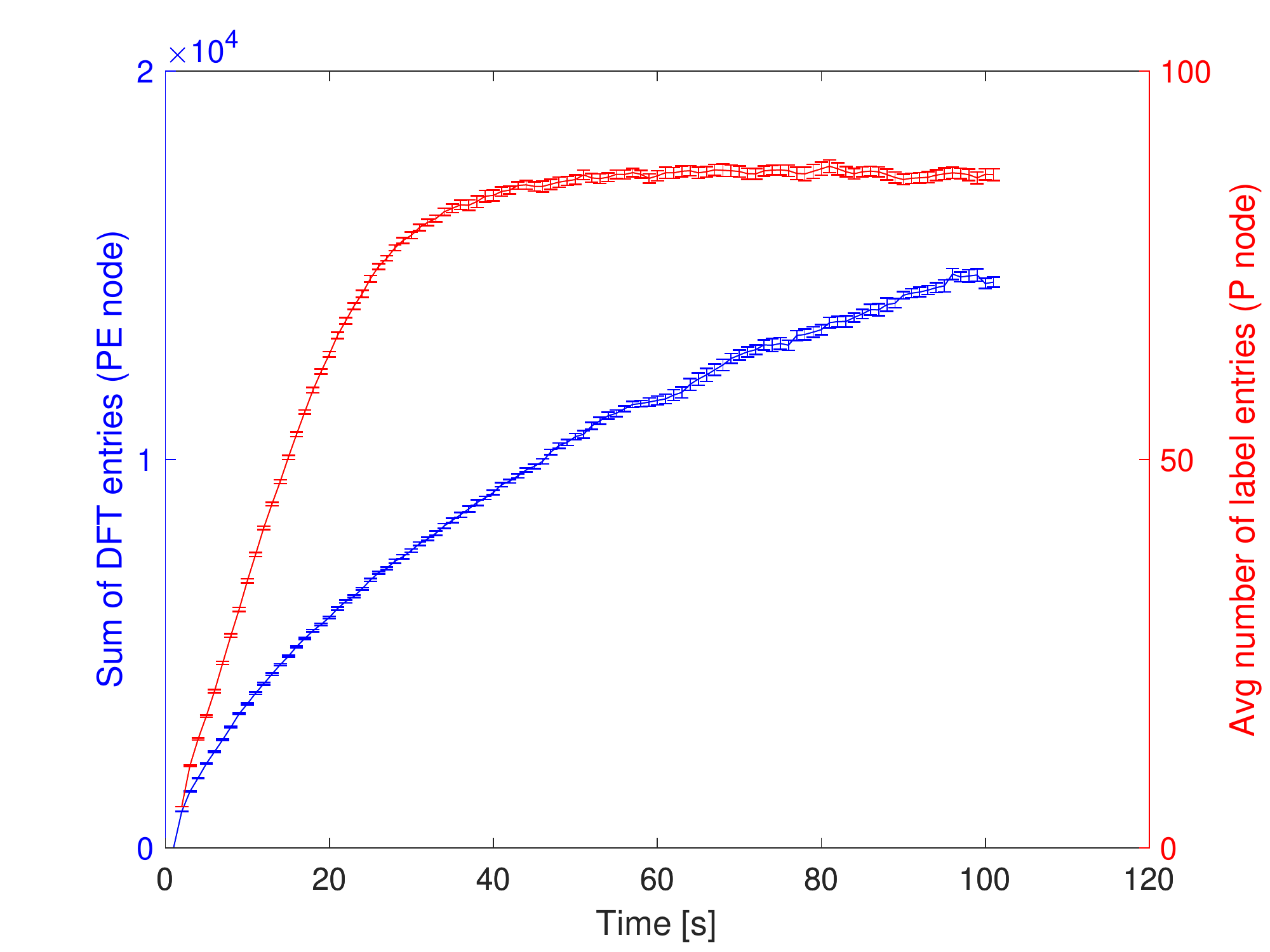}
    \caption{Average number of network flows and the used labels.} \label{fig:labels}
\end{figure}

During analysis of Tab.~\ref{tab:results}, we note that in the case of all congestion and warning threshold pairs, a notable increase of the total received data was observed ($\mathit{RxGain}$ of about 32\%). 
The main purpose to introduce our mechanism relates to the need for reduction of the number of flow entries in the core switches, i.e., P nodes. We can observe that the significant reduction has been obtained due to the flow aggregation procedure based on introduction of centrally managed MPLS label distribution performed by the SDN controller. Since all flows destined to DCNs attached to the particular PE node are represented by a single label, a large number of ingress flow entries from the edge of the network can be served by the same single label in the core. Then, the number of labels utilized by a single P node depends on the number of PE switches and the number of the used paths. The number of labels is sensitive to the statistics of flow life-times and the idle timeout value used by the garbage collector (in our simulations, the latter is set to 1~sec.). Since the network core forwards traffic from all the PE switches, it is instructive to compare the summarized number of flow entries in the network to the number of labels present in a single P switch. To see the impact of our mechanism, observe columns `Sum of DFT entries (PE)' and `Avg label entries (P)' in Tab.~\ref{tab:results}, where the difference of three orders of magnitude can be noticed. This result is well seen in Fig.~\ref{fig:labels}, where the time changes are shown. Despite the fact that the number of flows arriving to PE nodes (in blue) increases, the number of labels used by P nodes (in red) tends to stabilize. This observation confirms the high scalability achieved by the proposed mechanism.

Moreover, the indicator defined in Eq.~\eqref{eq:max-fri} proves potential and considerable scalability of our mechanism. Namely,
the $\mathit{maxFRI}$ illustrates the best achieved result for the considered network configuration and traffic conditions. In Tab.~\ref{tab:results}, the best achieved value of $\mathit{maxFRI}$ (99.55\%) is marked in bold. 

\begin{figure}[t]
    \centering
    \includegraphics[scale=.42,trim={100 0 100 0}]{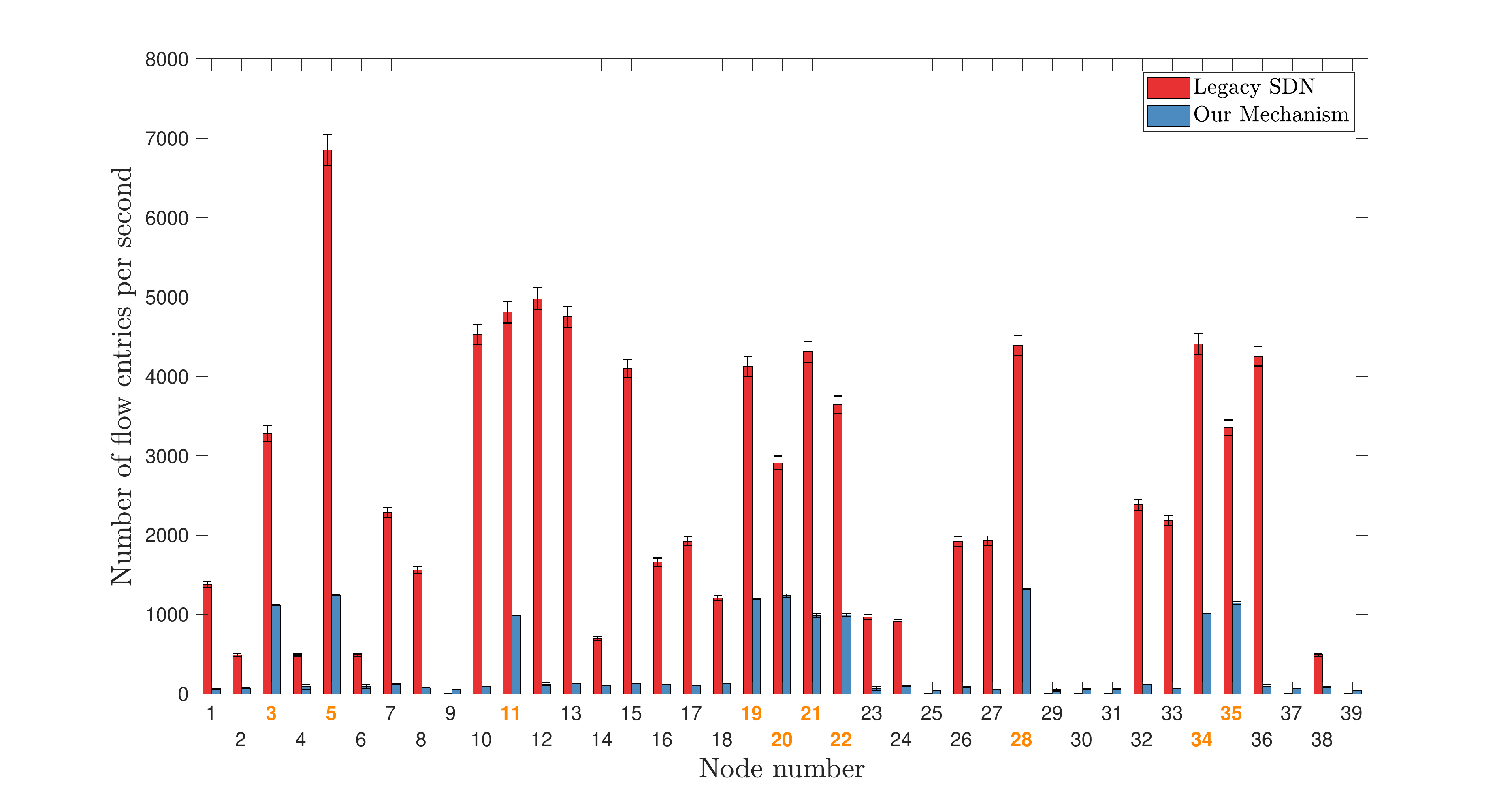} 
    \caption{Number of flow entries per node per second. PE node numbers are marked in orange.}\label{fig:flows}
\end{figure}

\begin{figure}[h!]
    \centering
    \includegraphics[scale=.7]{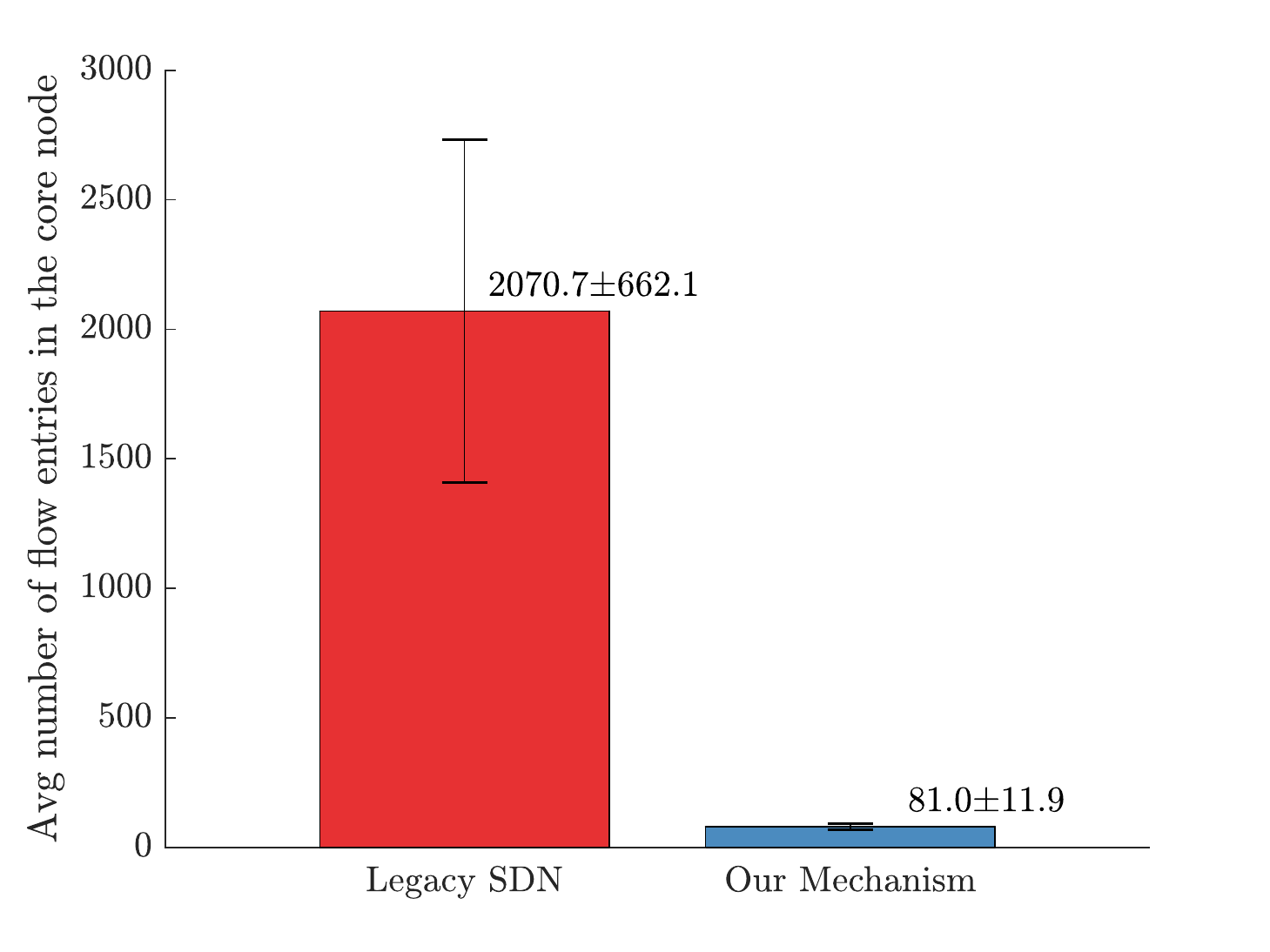} 
    \caption{Average number of flow entries in the core node.}\label{fig:avgflows}
\end{figure}

In Fig.~\ref{fig:flows}, we present the comparison of a number of flow entries in each node (per second) for the two scenarios: the `Legacy SDN' and our mechanism. As one can note, there are many more flow entries for the `Legacy SDN' case in comparison to our solution. This observation follows from the fact that when a network is heavily utilized a single flow is transmitted with lower throughput. In both scenarios flow inter-arrival time is the same. We use TCP flows, and they slow down if necessary, thus they are present for much longer periods in the network. That means that each single node has to maintain flows for a longer time. In result, the number of flow entries increases. In case of our mechanism, flows can achieve higher throughputs than for `Legacy SDN'. Consequently, the process of transmission is finished faster and flow entries in flow tables are maintained for shorter times. Thanks to multipath transmission, link congestions are avoided, and different flows between the same source-destination pairs may use distinct available paths. This way, network resources are utilized more efficiently. While comparing the `Legacy SDN' with our mechanism, one can see a significant reduction in the number of flow entries, especially for core nodes. This phenomenon can be observed in Fig.~\ref{fig:avgflows}, where the average number of flow entries for core nodes is shown.
In Fig.~\ref{fig:avgflows}, a considerable value of the confidence intervals for `Legacy SDN' follows from the fact that network resources are unevenly used, i.e., some core nodes are heavily utilized while others are used occasionally. Such the situation does not appear when our mechanism is applied, and then the network nodes are used uniformly. This fact can be easily seen in Fig.~\ref{fig:flows}, where the numbers of flow entries are almost the same for all the P nodes.

\begin{figure}[h!]
    \centering
    \includegraphics[scale=.42,trim={100 0 100 0}]{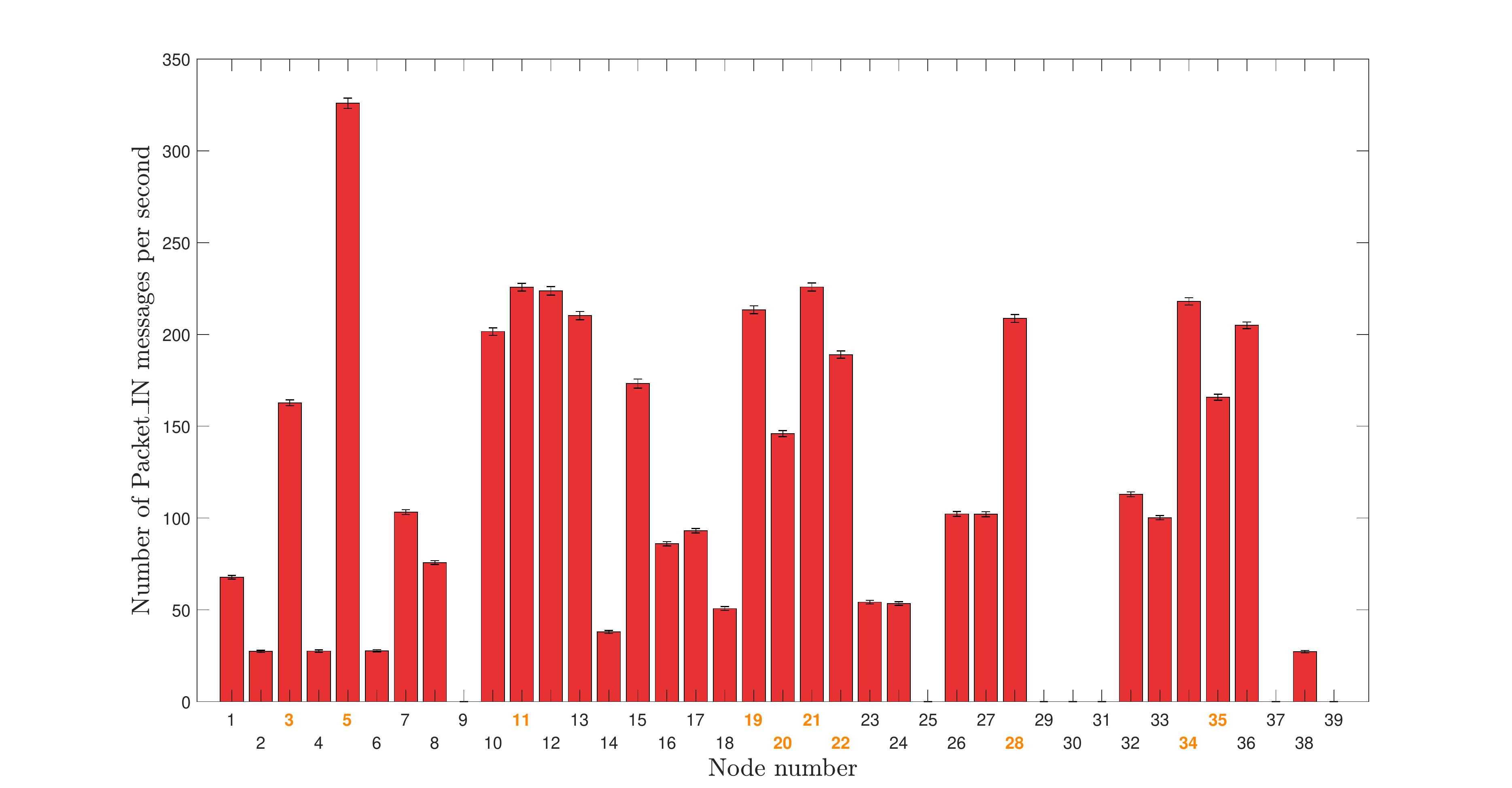} 
    \caption{Number of \texttt{Packet\_IN} messages per node per second for `Legacy SDN' scenario. PE node numbers are marked in orange.}\label{fig:pktin}
\end{figure}

Our mechanism significantly limits the communication between switches and an SDN controller. The controller only retrieves link statistics and updates aggregation entries (CFT in PE nodes, and label entries in P nodes). If we suppose that the controller collects link statistics and performs the reverse Dijkstra calculation every second (the worst case), the total number of exchanged messages between the controller and the nodes per second is equal to $117 = 2 \times 39 + 39$ (the request and response for statistics from 39 nodes summed with distribution of flow entries with labels match using bundle messages). The number of \texttt{Packet\_IN} messages per second for `Legacy SDN' is presented in Fig.~\ref{fig:pktin} for each node separately. Some nodes do not belong to any path connecting source-destination nodes, therefore the number of \texttt{Packet\_IN} messages equals zero for them. The same reasoning applies to the situation presented in Fig.~\ref{fig:flows}, where the number of flow entries for such nodes equals zero as well.

In Tab.~\ref{tab:comparison}, we present some efficiency indicators for the simulated mechanisms. We deal with the TCP traffic only, thus the total amount of the transmitted data strongly depends on the amount of bandwidth available in the network. When our mechanism is used, a larger number of paths to each destination is available, therefore the TCP sources are able to send extended data volumes. This stems from the fact that TCP window sizes can be steadily increased. We are able to observe the increased overall transmission throughput in comparison to the `Legacy SDN' case (`Avg Tput' column). We want to stress that in all the simulations of the `Legacy SDN' case, we do not implement the delay related to flow installation (including \texttt{Packet\_IN} messages generation, processing and insertion of the resulting flow in the switches). While looking at the total number of OpenFlow messages served by the SDN controller per second (the `No. of OpenFlow messages' column), we can conclude that this delay can significantly influence the throughput. Moreover, it is likely that packets belonging to a single flow can arrive with such a high frequency that the installation of a forwarding rule in a switch takes place after many packets from the same flow arrive. This results in triggering many unnecessary \texttt{Packet\_IN} messages, causing further overload of the controller. Additionally, we want to notice that in our simulations, we consider the situation that only a single \texttt{Packet\_IN} message is generated for each separate flow. Therefore, in real networks the throughput for `Legacy SDN' is expected to be even lower than the one presented here. In our simulations, the total number of OpenFlow messages is equal to a doubled number of \texttt{Packet\_IN} messages ($2 \times 4238.37$). We observe as high as 98\% reduction of the number of OpenFlow messages exchanged between the controller and switches when our mechanism is compared to the `Legacy SDN' case. We want to notice that for this calculation we used the maximum number of exchanged messages for our mechanism (the worst case), and the number of OpenFlow messages obtained from `Legacy SDN' simulations, where all messages are processed instantaneously. The achieved reduction of flow entries in the core nodes (defined by Eq.~\eqref{eq:fri}) in comparison to `Legacy SDN' is summarised in the `$\mathit{CFRI}$' column of Tab.~\ref{tab:comparison}.

\begin{table}
\centering
\caption{Comparison of the mechanisms}
\label{tab:comparison}
\begin{tabular}{c|c|c|c|}
\cline{2-4}
                                    & Avg Tput              & No. of OpenFlow  & $\mathit{CFRI}$ \\ 
                                    & [Mbps]                & messages         & [\%]   \\ \hline
\multicolumn{1}{|c|}{Legacy SDN} & 995.3$\pm$7.4         & 8476.74$\pm$33.57          & 95.78$\pm$0.06     \\ \hline
\multicolumn{1}{|c|}{Proposed mechanism} & 1319.8$\pm$8.7        & $\leq$ 117                          & N/A                 \\ \hline
\end{tabular}
\end{table}

%% file: sections/related.tex
\section{Related Work} \label{sec:related}

Some proposals of flow table reduction have appeared in literature. The authors of FAMTAR~\cite{famtar} (decentralized mechanism, not SDN-based) introduced aggregation procedure limiting the number of flows in core routers~\cite{famtar-aggr}. Separate flows following the same routing path are aggregated by marking them with a tag. The tag represents a sequence of nodes on an OSPF path connecting entry and exit nodes of a domain using FAMTAR. Similarly to the majority of aggregation mechanisms, nodes are also separated into two classes: edge and core. Edge nodes use three tables, namely standard Routing Table (RT), Flow Forwarding Table (FFT) and Aggregation Table (AT). Core routers posess RT and AT. When a new packet arrives to an edge of the domain, the edge node does not have an entry in FFT. It consults the RT and AT to build a new FFT entry containing a flow identity, and a tag (from AT), as well as an output interface (from RT). Then, packets belonging to existing flows are processed in accordance with FFT. When a packet reaches a core node, it looks up its AT. If match of the tag is not found, the router looks up its RT for the destination. Next, it creates an AT entry containing the received tag identity and the output interface (retrieved from RT).

The number of FATMAR's FFT entries and the number of DFT entries (in our mechanism) for any considered scenario are the same. The both mechanisms significantly differ in number of entries stored at the core nodes. Let us consider a network containing 100 edge nodes, all being entrances and exits of a domain (sources and destinations of a traffic). Suppose that there are no congested links in the network. For the FAMTAR solution, each connection between edge nodes is marked with different tag. This gives the total number of $9900 = 99 \times 100$ (99 destinations from each 100 sources) tags in the whole network. This number represents the number of flow entries a single core node has to handle in the worst case, i.e., this core node has to process communication with all edge nodes. In our case, each connection from any edge node to a particular exit node is tagged with the same single global MPLS label. This results in the total number of 100 tags in the whole network. This number is also obtained in the worst case. When multipath transmission is considered, the both mechanisms recalculate routing in the network. Supposing the worst case scenario, where all the existing flows are still present on the previous paths (before the path recalculation takes place), all the old tags have to be maintained. For the new flows new tags have to be allocated. Therefore, a single change of routing induces increase of a number of used tags to $19880 = 2 \times 9900$ for FAMTAR. When we consider our mechanism, the core nodes have to maintain only $200 = 2 \times 100$ tags (labels). This way, our mechanism is much more scalable.

On the other hand, the authors of~\cite{2mpls} introduce partitioning of a network into regions and assume use of two MPLS tags to reduce a number of flow table entries. To compare this mechanism to FAMTAR and our mechanism let us consider previously used scenario again. To be consistent with our terminology, we use the `edge' notion for the nodes where host networks are connected. When a network posessing 100 edge nodes is divided into 10 regions, each with 10 edge nodes, the maximum number of flow entries in a core node is 20 for mechanism proposed in~\cite{2mpls}. Such the number is valid when only one node is a communication point for other regions. This mechanism offers a better flow scalability than our proposition and FAMTAR. In~\cite{2mpls}, the authors use an optimization task for selection of a number of regions. 10 is the optimal value for regions, resulting in a lower number of flow table entries for the proposed scenario. When one decides to change the number of regions, the average number of flows increases according to the following rule: $R+E/R$, where $R$ is a number of regions and $E$ is a number of edge routers. However, the authors do not offer any congestion avoidance procedure. In our opinion, use of a single node as an entrance to each region increases probability of a congestion. We cannot qualitatively compare this mechanism with our solution because paper~\cite{2mpls} does not introduce any multipath transmission method. The next advantage of our mechanism present contrary to~\cite{2mpls} is a lack of communication between switches and a controller for every new flow installation (excluding the extension presented is Subsection~\ref{sec:extension}). The compared mechanism~\cite{2mpls} requires a reactive installation of each new flow in edge nodes.

In~\cite{dynamic-agg}, the authors propose a dynamic flow aggregation based on VLAN tagging. The basis for the aggregation procedure is a common path that many flows traverse in the same time. Since authors of~\cite{dynamic-agg} give the performance evaluation based on a ring, a tree and a ring of trees topologies, it is hard to compare their solution with our universal mechanism. They obtain 48\% reduction of flow entries in the core of a network, comparing to 96\% in our case. In this solution, a controller has to maintain a huge database storing information about all the running flows. We perceive this as a week point. When a packet belonging to a new flow arrives to a switch, always a \texttt{Packet\_IN} message is generated independently whether aggregate exists or not. Contrary to our approach, this solution creates enormous communication burden between network nodes and the controller.

In~\cite{curtis2011}, Curtis \emph{et al.} introduce modification of OpenFlow protocol, known as DevoFlow. In this solution, similarly to our proposal, a switch is able to insert some flow rules into a flow table on its own. The authors distinguish two classes of flows based on measurements: mice and elephants. The former can be installed by a switch itself using proper wildcards, while the latter require communication with a controller. When a given flow is recognized as an elephant flow, a switch consults the controller. It finds the least occupied path to the destination and re-routes this flow by installing proper flow rules
in switches on the path found. Communication between network nodes and the controller is minimized, but still may be significant (depending on the distribution of elephant flows). Curtis \emph{et al.} explore also a multipath transmission enabling congestion avoidance. In contrast to our multipath approach, re-routing of elephant flows may results in packet reordering. 

Wang \emph{et al.}~\cite{wang2017} focus on minimization of the delay between pairs of nodes using precomputed multiple paths. Nevertheless, the decrease of latency is obtained at a cost of increase in a number of flow entries at switches. Our approach avoids this trade-off. The solution presented in~\cite{lin2016} concentrates on minimization of communication between the controller and switches. A separate pool of tags for marking different paths (one per a source-destination pair) is used. Contrary to our mechanism, this proposal does not implement dynamic reaction to congestions and each new flow must be processed by the controller, while our solution offers a better flow aggregation in the core network and significantly reduces number of messages exchanged between the SDN controller and switches. 

%% file: sections/conclusion.tex
\section{Conclusions} \label{sec:conclusion}
To improve the network utilization and efficiency of SDN-based forwarding, we propose the mechanism for flow aggregation accompanied with multipath transmission. The performed experiments show that the application of the mechanism results in 96\% reduction of flow entries in core nodes and 98\% reduction of OpenFlow messages. We also observe that the overall network traffic increases by about 32\%. The significant reduction in a number of flow entries in the core of the network is obtained due to the flow aggregation procedure. The procedure is enabled by introduction of centrally managed MPLS label distribution. The distribution is performed by an SDN controller without application of legacy signalling protocols. The increase of the network traffic has been obtained due to application of multipath transmission. When a potential link overload is detected, the mechanism dynamically finds new paths for new flows. Beside the mentioned advantages, our mechanism does not involve new protocols, but uses only simple modifications of the existing solutions. That is, a modified switch is able to install new flow rules on its own. Thanks to that communication between the SDN controller and network nodes is minimized. Moreover, the mechanism may be deployed incrementally using legacy MPLS nodes in core of the network.

\section*{Acknowledgment}

This research was carried out with the supercomputer ``Deszno'' purchased thanks to the financial support of the European Regional Development Fund in the framework of the Polish Innovation Economy Operational Program (contract no.\ POIG.\ 02.01.00-12-023/08).

This research also was supported in part by PL-Grid Infrastructure. 

%% file: arXiv_mpls_sdn.bbl
\begin{thebibliography}{10}
\providecommand{\url}[1]{#1}
\csname url@samestyle\endcsname
\providecommand{\newblock}{\relax}
\providecommand{\bibinfo}[2]{#2}
\providecommand{\BIBentrySTDinterwordspacing}{\spaceskip=0pt\relax}
\providecommand{\BIBentryALTinterwordstretchfactor}{4}
\providecommand{\BIBentryALTinterwordspacing}{\spaceskip=\fontdimen2\font plus
\BIBentryALTinterwordstretchfactor\fontdimen3\font minus
  \fontdimen4\font\relax}
\providecommand{\BIBforeignlanguage}[2]{{%
\expandafter\ifx\csname l@#1\endcsname\relax
\typeout{** WARNING: IEEEtran.bst: No hyphenation pattern has been}%
\typeout{** loaded for the language `#1'. Using the pattern for}%
\typeout{** the default language instead.}%
\else
\language=\csname l@#1\endcsname
\fi
#2}}
\providecommand{\BIBdecl}{\relax}
\BIBdecl

\bibitem{fortz2002}
B.~Fortz and M.~Thorup, ``{Optimizing OSPF/IS-IS Weights in a Changing
  World},'' \emph{IEEE Journal on Selected Areas in Communications}, vol.~20,
  no.~4, pp. 756--767, May 2002.

\bibitem{multipath-survey1}
J.~Dom\.{z}a\l, Z.~Duli\'{n}ski, M.~Kantor, J.~Rzk{a}sa, R.~Stankiewicz,
  K.~Wajda, and R.~W\'{o}jcik, ``{A Survey on Methods to Provide Multipath
  Transmission in Wired Packet Networks},'' \emph{Computer Networks}, vol.~77,
  pp. 18 -- 41, 2015.

\bibitem{multipath-survey2}
R.~W\'{o}jcik, J.~Dom\.{z}a\l, Z.~Duli\'{n}ski, G.~Rzym, A.~Kamisi\'{n}ski,
  P.~Gaw\l{}owicz, P.~Jurkiewicz, J.~Rz\k{a}sa, R.~Stankiewicz, and K.~Wajda,
  ``{A Survey on Methods to Provide Interdomain Multipath Transmissions},''
  \emph{Computer Networks}, vol. 108, no.~C, pp. 233--259, Oct. 2016.

\bibitem{rfc3031}
E.~Rosen, A.~Viswanathan, and R.~Callon, ``{Multiprotocol Label Switching
  Architecture},'' \emph{IETF RFC 3031}, 2001.

\bibitem{sdnsurvey}
W.~Xia, Y.~Wen, C.~H. Foh, D.~Niyato, and H.~Xie, ``{A Survey on
  Software-Defined Networking},'' \emph{IEEE Communications Surveys and
  Tutorials}, vol.~17, no.~1, pp. 27--51, 2015.

\bibitem{sdnwhitepaper}
``{Software-Defined Networking: The New Norm for Networks},'' \emph{Open
  Networking Foundation whitepaper}, 2012.

\bibitem{openflow}
N.~McKeown, T.~Anderson, H.~Balakrishnan, G.~Parulkar, L.~Peterson, J.~Rexford,
  S.~Shenker, and J.~Turner, ``{OpenFlow: Enabling Innovation in Campus
  Networks},'' \emph{SIGCOMM Comput. Commun. Rev.}, vol.~38, no.~2, pp. 69--74,
  Mar. 2008.

\bibitem{openflowspec}
``{OpenFlow Switch Specification v1.5.1},'' \emph{Open Networking Foundation
  specification}, 2015.

\bibitem{curtis2011}
A.~R. Curtis, J.~C. Mogul, J.~Tourrilhes, P.~Yalagandula, P.~Sharma, and
  S.~Banerjee, ``{DevoFlow: Scaling Flow Management for High-performance
  Networks},'' \emph{SIGCOMM Comput. Commun. Rev.}, vol.~41, no.~4, pp.
  254--265, Aug. 2011.

\bibitem{khalili2016}
R.~Khalili, W.~Y. Poe, Z.~Despotovic, and A.~Hecker, ``{Reducing State of
  OpenFlow Switches in Mobile Core Networks by Flow Rule Aggregation},'' in
  \emph{Proc. 2016 25\textsuperscript{th} International Conference on Computer
  Communication and Networks (ICCCN)}, Aug 2016, pp. 1--9.

\bibitem{chuang2016}
C.~C. Chuang, Y.~J. Yu, A.~C. Pang, and G.~Y. Chen, ``{Minimization of TCAM
  Usage for SDN Scalability in Wireless Data Centers},'' in \emph{Proc. 2016
  IEEE Global Communications Conference GLOBECOM}, Dec. 2016, pp. 1--7.

\bibitem{commag-scalability}
S.~H. Yeganeh, A.~Tootoonchian, and Y.~Ganjali, ``{On Scalability of
  Software-defined Networking},'' \emph{IEEE Communications Magazine}, vol.~51,
  no.~2, pp. 136--141, February 2013.

\bibitem{nox}
A.~Tavakoli, M.~Casado, T.~Koponen, and S.~Shenker, ``{Applying NOX to the
  Datacenter},'' in \emph{Proc. Workshop on Hot Topics in Networks
  HotNets-VIII}, Oct. 2009.

\bibitem{controller-performance}
A.~Tootoonchian, S.~Gorbunov, Y.~Ganjali, M.~Casado, and R.~Sherwood, ``{On
  Controller Performance in Software-defined Networks},'' in \emph{Proc.
  2\textsuperscript{nd} USENIX Conference on Hot Topics in Management of
  Internet, Cloud, and Enterprise Networks and Services Hot-ICE'12}, Apr. 2012,
  pp. 10--10.

\bibitem{benson}
T.~Benson, A.~Akella, and D.~A. Maltz, ``{Network Traffic Characteristics of
  Data Centers in the Wild},'' in \emph{Proc. 10\textsuperscript{th} ACM
  SIGCOMM Conference on Internet Measurement IMC'10}.\hskip 1em plus 0.5em
  minus 0.4em\relax New York, NY, USA: ACM, 2010, pp. 267--280.

\bibitem{Kandula}
S.~Kandula, S.~Sengupta, A.~Greenberg, P.~Patel, and R.~Chaiken, ``{The Nature
  of Data Center Traffic: Measurements \& Analysis},'' in \emph{Proceedings of
  the 9th ACM SIGCOMM Conference on Internet Measurement}, ser. IMC '09.\hskip
  1em plus 0.5em minus 0.4em\relax New York, NY, USA: ACM, 2009, pp. 202--208.

\bibitem{sr-draft}
C.~Filsfils, S.~Previdi, L.~Ginsberg, B.~Decraene, S.~Litkowski, and R.~Shakir,
  ``{Segment Routing Architecture},''
  \emph{draft-ietf-spring-segment-routing-15}, 2018.

\bibitem{tcam-razor}
A.~X. Liu, C.~R. Meiners, and E.~Torng, ``{TCAM Razor: A Systematic Approach
  Towards Minimizing Packet Classifiers in TCAMs},'' \emph{IEEE/ACM
  Transactions on Networking}, vol.~18, no.~2, pp. 490--500, April 2010.

\bibitem{Katta2014}
N.~Katta, O.~Alipourfard, J.~Rexford, and D.~Walker, ``{Infinite CacheFlow in
  Software-defined Networks},'' in \emph{Proceedings of the Third Workshop on
  Hot Topics in Software Defined Networking HotSDN '14}.\hskip 1em plus 0.5em
  minus 0.4em\relax New York, NY, USA: ACM, 2014, pp. 175--180.

\bibitem{rfc2992}
C.~Hopps, ``{Analysis of an Equal-Cost Multi-Path Algorithm},'' \emph{IETF RFC
  2992}, 2000.

\bibitem{rfc7868}
D.~Savage, J.~Ng, S.~Moore, D.~Slice, P.~Paluch, and R.~White, ``{Cisco's
  Enhanced Interior Gateway Routing Protocol (EIGRP)},'' \emph{IETF RFC 7868},
  2016.

\bibitem{latency-sdn}
K.~He, J.~Khalid, S.~Das, A.~Gember-Jacobson, C.~Prakash, A.~Akella, L.~E. Li,
  and M.~Thottan, ``{Latency in Software Defined Networks: Measurements and
  Mitigation Techniques},'' \emph{SIGMETRICS Perform. Eval. Rev.}, vol.~43,
  no.~1, pp. 435--436, June 2015.

\bibitem{measuring-latency}
K.~He, J.~Khalid, A.~Gember-Jacobson, S.~Das, C.~Prakash, A.~Akella, L.~E. Li,
  and M.~Thottan, ``{Measuring Control Plane Latency in SDN-enabled
  Switches},'' in \emph{Proceedings of the 1st ACM SIGCOMM Symposium on
  Software Defined Networking Research SOSR'15}, June 2015.

\bibitem{ns3}
\BIBentryALTinterwordspacing
``{The ns-3 Discrete-Event Network Simulator},'' 2018. [Online]. Available:
  \url{https://www.nsnam.org}
\BIBentrySTDinterwordspacing

\bibitem{us-net}
S.~Orlowski, R.~Wess\"{a}ly, M.~Pi\'{o}ro, and A.~Tomaszewski, ``{SNDlib 1.0
  --- Survivable Network Design Library},'' \emph{Networks}, vol.~55, no.~3,
  pp. 276--286, May 2010.

\bibitem{famtar}
R.~W\'{o}jcik, J.~Dom\.{z}a\l, and Z.~Duli\'{n}ski, ``{Flow-Aware
  Multi-Topology Adaptive Routing},'' \emph{IEEE Communications Letters},
  vol.~18, no.~9, pp. 1539--1542, Sept. 2014.

\bibitem{famtar-aggr}
J.~Dom\.{z}a\l, P.~Jurkiewicz, P.~Gaw\l{}owicz, and R.~W\'{o}jcik, ``{Flow
  Aggregation Mechanism for Flow-Aware Multi-Topology Adaptive Routing},''
  \emph{IEEE Communications Letters}, vol.~21, no.~12, pp. 2582--2585, Dec.
  2017.

\bibitem{2mpls}
N.~Kitsuwan, S.~Ba, E.~Oki, T.~Kurimoto, and S.~Urushidani, ``{Flows Reduction
  Scheme Using Two MPLS Tags in Software-Defined Network},'' \emph{IEEE
  Access}, vol.~5, pp. 14\,626--14\,637, 2017.

\bibitem{dynamic-agg}
A.~Mimidis, C.~Caba, and J.~Soler, ``{Dynamic Aggregation of Traffic Flows in
  SDN: Applied to Backhaul Networks},'' in \emph{{2016 IEEE NetSoft Conference
  and Workshops (NetSoft)}}, June 2016, pp. 136--140.

\bibitem{wang2017}
Y.-C. Wang, Y.-D. Lin, and G.-Y. Chang, ``{SDN-based Dynamic Multipath
  Forwarding for Inter-data Center Networking},'' in \emph{Proc. 2017 IEEE
  International Symposium on Local and Metropolitan Area Networks LANMAN}, June
  2017, pp. 1--3.

\bibitem{lin2016}
W.~Lin, Y.~Niu, X.~Zhang, L.~Wei, and C.~Zhang, ``{Using Path Label Routing in
  Wide Area Software-Defined Networks with OpenFlow},'' in \emph{Proc. 2016
  International Conference on Networking and Network Applications NaNA}, July
  2016, pp. 149--154.

\end{thebibliography}
